\documentclass[usegraphicx,usenatbib]{mn2e} 

\usepackage{url}
\usepackage[breaklinks=true]{hyperref}
\usepackage{twoopt}
\usepackage[english]{babel}          

\usepackage{natbib}
\bibpunct{(}{)}{;}{a}{}{,} 

\usepackage[utf8x]{inputenc}
\usepackage[normalem]{ulem}
\usepackage{siunitx}
\usepackage{amsmath, amssymb}
\usepackage[T1]{fontenc} 
\usepackage{aecompl} 

\usepackage[farskip=0pt]{subfig}

\usepackage{multirow,bigstrut,ctable}
\usepackage[normalem]{ulem}
\usepackage{rotating}
\usepackage[varg]{txfonts} 
\usepackage{threeparttable}

\def \kms         {km$\,$s$^{-1}$}

\def \deg         {\text{$^{\circ}$}}
\def \arcmin      {\text{$^\prime$}}
\def \arcsec      {\text{$^{\prime\prime}$}}
\def \hour        {$^{\mathrm{h}}$}
\def \min         {$^{\mathrm{m}}$}
\def \sec         {$^{\mathrm{s}}$}
\def \mjybeam     {mJy\,beam$^{-1}$}
\def \mujybeam    {$\mu$Jy\,beam$^{-1}$}

\def \mach	  {\mathcal{M}}

\newcommand{\hms}[3]{{#1}\hour{#2}\min{#3}\sec}
\newcommand{\dms}[3]{{#1}\deg{#2}\arcmin{#3}\arcsec}
\newcommand{\beam}[2]{{#1}\arcsec$\times${#2}\arcsec}

\newcommand{\Msun}{\text{$\rm M_\odot$}}
\newcommand{\ion}[2]{#1{\small\uppercase\expandafter{\romannumeral#2}}\relax}

\def \target {PSZ1 G108.18-11.53}

\title[The radio relic system in PSZ1 G108.18-11.53]
      {A powerful double radio relic system discovered in PSZ1 G108.18-11.53: evidence for a shock with non-uniform Mach number?}

\author[F.~de~Gasperin et~al.]{F. de Gasperin$^{1}$, H. T. Intema$^{3}$, R. J. van Weeren$^{2}$, W. A. Dawson$^{4}$,
\newauthor N. Golovich$^{5}$, D. Wittman$^{5}$, A. Bonafede$^{1}$, M. Br\"uggen$^{1}$
\\
\\
$^{1}$ Universit\"at Hamburg, Hamburger Sternwarte, Gojenbergsweg 112, D-21029, Hamburg, Germany\\
$^{2}$ Harvard-Smithsonian Center for Astrophysics, 60 Garden Street, Cambridge, MA 02138, USA\\
$^{3}$ National Radio Astronomy Observatory, 1003 Lopezville Road, Socorro, NM 87801-0387, USA\\
$^{4}$ Lawrence Livermore National Lab, 7000 East Avenue, Livermore, CA 94550, USA\\
$^{5}$ University of California, One Shields Avenue, Davis, CA 95616, USA\\
}      
      
\begin{document}

\date{LLNL-JRNL-661314-DRAFT}
\pagerange{\pageref{firstpage}--\pageref{lastpage}} \pubyear{2013}
\maketitle

\label{firstpage}

\begin{abstract}
Diffuse radio emission in the form of radio halos and relics has been found in a number of merging galaxy clusters. These structures indicate that shock and turbulence associated with the merger accelerate electrons to relativistic energies.
We report the discovery of a radio relic + radio halo system in \target{} ($z=0.335$). This cluster hosts the second most powerful double radio relic system ever discovered.
We observed \target{} with the Giant Meterwave Radio Telescope (GMRT) and the Westerbork Synthesis Radio Telescope (WSRT). We obtained radio maps at 147, 323, 607 and 1380 MHz. We also observed the cluster with the Keck telescope, obtaining the spectroscopic redshift for 42 cluster members.
From the injection index we obtained the Mach number of the shocks generating the two radio relics. For the southern shock we found $\mach = 2.33^{+0.19}_{-0.26}$, while the northern shock Mach number goes from $\mach = 2.20^{+0.07}_{-0.14}$ in the north part down to $\mach = 2.00^{+0.03}_{-0.08}$ in the southern region. If the relation between the injection index and the Mach number predicted by diffusive shock acceleration (DSA) theory holds, this is the first observational evidence for a gradient in the Mach number along a galaxy cluster merger shock.
\end{abstract}


\section{Introduction}

Cosmic structure forms hierarchically, with smaller structures merging to form bigger ones. On the largest scales, clusters of galaxies merge releasing energies of the order of $10^{64}$ erg on time scales of 1--2 Gyr \citep[e.g.][]{Hoeft2008}. During such merger events, large-scale shock waves with moderate Mach numbers are created in the intracluster medium (ICM). Shocks are collisionless features whose interactions in the hot plasma are mediated by electromagnetic fields. They act on the ICM accelerating a fraction of the thermal distribution of particles transforming them into nonthermal populations of cosmic rays (CRs) through DSA \citep[e.g.][]{Ensslin1998}. 

Uncertainties remain on the mechanism required to pre-accelerate the electrons from thermal to supra-thermal energies before the injection in the DSA process, especially at low Mach numbers and in high beta plasmas \citep[e.g.][]{Kang2014, Vazza2014}. Recent developments in the field have come from Particle-in-Cell (PIC) simulations that demonstrated that self-excitation of plasma waves via various instabilities may be crucial in the injection process \citep{Caprioli2014,Guo2014}. A second mechanism that is expected to accelerate particles is turbulence amplified by cluster mergers \citep[see][and reference therein]{Brunetti2014}. Evidence for the production of CRs in merging galaxy clusters by these two mechanisms has been found primarily in the form of radio relics and halos respectively.

\textit{Radio relics} are polarized, low surface brightness ($\simeq10^{-6}$ Jy arcsec$^{-2}$ at 1.4 GHz), steep-spectrum ($\alpha < -1$, with $S_{\nu} \propto \nu^\alpha$) sources that are typically located near the periphery of the cluster that underwent a recent merger. Relics are elongated and can reach sizes of 1-2 Mpc. The most likely scenario is that the synchrotron emitting electrons are (re)accelerated by outward moving shocks to a Lorentz factor of $\gamma_e \sim 10^3 - 10^5$ \citep[e.g.][]{Blandford1987}.

\textit{Radio halos} are Mpc-size sources generated by radio-emitting plasma permeating the central parts of disturbed clusters. One of the potential mechanisms to form radio halos is the acceleration of relativistic particles by turbulence, generated in the ICM by cluster-cluster mergers \citep{Brunetti2001,Petrosian2001}. Alternatively, the energetic electrons may be secondary products of proton-proton collisions \citep[e.g.][]{Dennison1980}.

\subsection{Double radio relics}

Models predict that giant radio relics typically occur along the cluster's merging axis and their orientation should be perpendicular to the axis itself. Simulations of binary mergers \citep{vanWeeren2012} show that only mergers whose axes are fairly aligned with the plane of the sky can develop a double system of radio relics. A double radio relic system has a major advantage: the approximate knowledge of the merger axis orientation makes it easier to correct for projection effects and obtain the relic's real distance from the cluster centre. Furthermore, when the merger axis is aligned with the plane of the sky, X-ray and optical observations are easier to interpret and cross-match with radio data. Only 15 other systems with double radio relics have been discovered so far \citep{deGasperin2014c}.

In this paper we report the discovery of a new double radio relic + radio halo system in \target{} ($z\approx0.335$). The \cite{PlanckCollaboration2013} recently discovered this cluster by detecting its Sunyaev-Zel'dovich (SZ) signal. \target{} hosts one of the most powerful double radio relic system known to date and its relics have a very well defined morphology (non-disrupted arc-like structure, no strong blending with other sources, high surface brightness). These characteristics allowed us to perform a detailed spectral analysis which led to the first observational evidence of a non-uniform Mach number distribution along the shock that generated one of the relics.

In Sec.~\ref{sec:observations} we present new radio and optical data. In Sec.~\ref{sec:target} we present the global properties of the hosting cluster and of the radio relic (Sec.~\ref{sec:relics}) and halo (Sec.~\ref{sec:halo}) system. In Sec.~\ref{sec:spanalysis} we analyse the radio spectra of the sources. Discussion and conclusions are in Sec.~\ref{sec:discussion} and Sec.~\ref{sec:conclusions}, respectively.
Throughout the paper we adopt a fiducial $\Lambda$CDM cosmology with $H_0 = 70\rm\ km\ s^{-1}\ Mpc^{-1}$, $\Omega_m = 0.3$ and $\Omega_\Lambda = 0.7$. At the redshift of the target ($z\approx0.335$) 1\arcsec = 4.8 kpc. All regressions are made with an ordinary least squares algorithm that takes into account errors only in the dependant variable. Unless otherwise specified errors are at $1\sigma$.

\section{Observations}
\label{sec:observations}

The cluster was observed with the Westerbork Radio Synthesis Telescope (WSRT) at 21 cm (see Sec.~\ref{sec:WSRT}) and subsequently with the Giant Meterwave Radio Telescope (GMRT) at three separate frequencies to enable spectral analysis (see Sec.~\ref{sec:GMRT}). A summary of the radio observations is listed in Table~\ref{tab:observations}. A preliminary optical follow-up was also made using the W.M.Keck telescope (see Sec.~\ref{sec:opt}).

\begin{table}
\centering
\begin{threeparttable}
\begin{tabular}{llllll}
          & Freq.     & Obs. & Length & Resolution & rms\\
          & (MHz)     & date & (h)    &            & (\mujybeam) \\
\hline
GMRT & 147 & 11 Jun 14 & 8 & \beam{22}{13} & 1600 \\
GMRT & 323 & 02 Jun 14 & 8 & \beam{8}{6} & 86 \\
GMRT & 607 & 06 Jun 14 & 8 & \beam{4}{4} & 43 \\
WSRT & 1380 & 02 Jan 14 & 12 & \beam{17}{13} & 37 \\
\end{tabular}
\end{threeparttable}
\caption{Radio observations}\label{tab:observations}
\end{table}

\subsection{Radio: WSRT}
\label{sec:WSRT}
The cluster was observed with the WSRT on January 2nd, 2014, for about 12 h using the default 21~cm set-up. Due to telescope upgrades, only 8 antennas participated in the observation. A total bandwidth of 160~MHz was recorded, spread over eight spectral windows of 20~MHz in bandwidth. Each spectral window was further subdivided into 64 frequency channels. All four linear polarization products were recorded. The calibrators 3C286 and 3C147 were observed at the start and end of the main observing run, respectively.

\begin{figure*}
\centering
\includegraphics[width=\columnwidth]{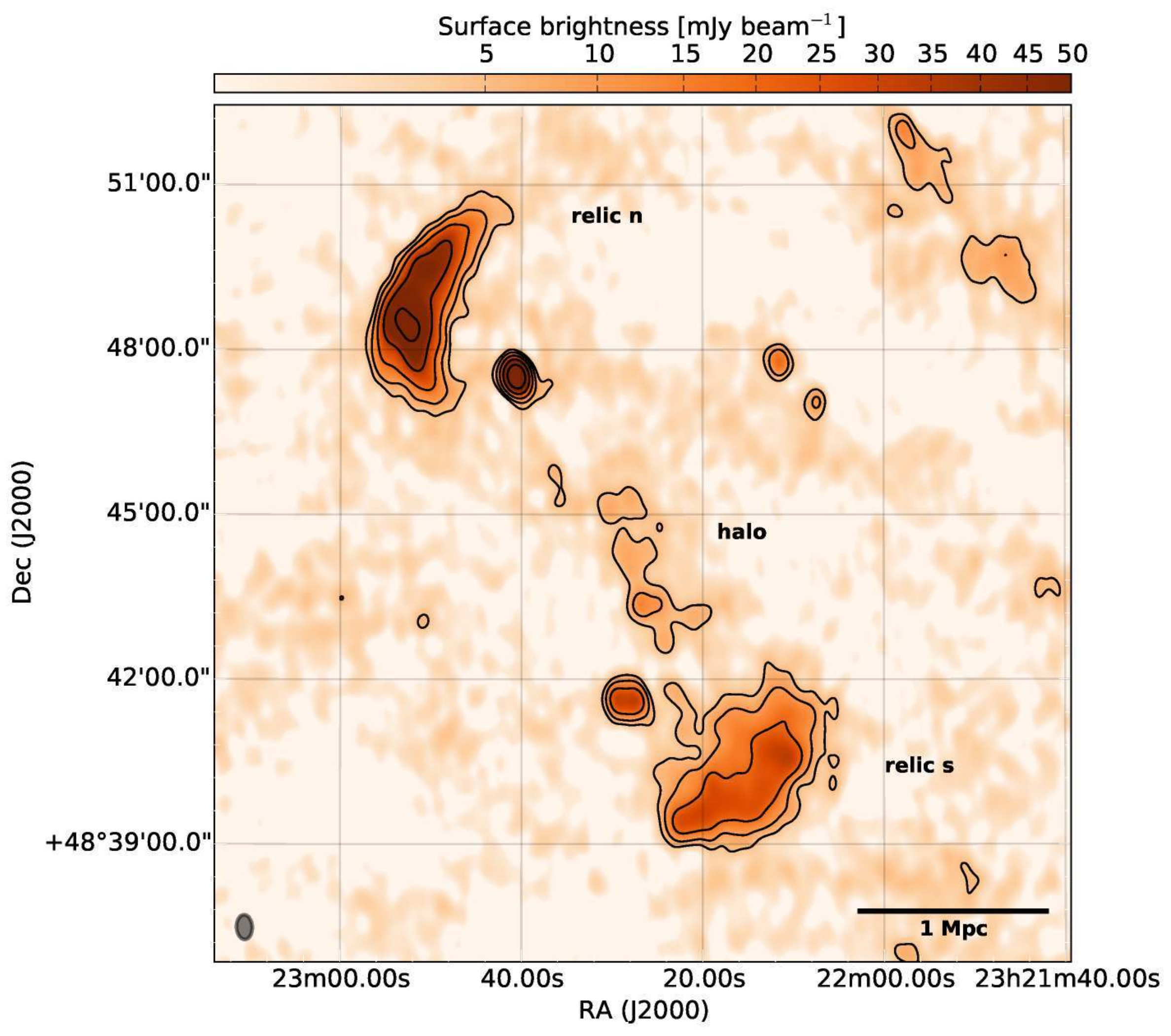}
\includegraphics[width=\columnwidth]{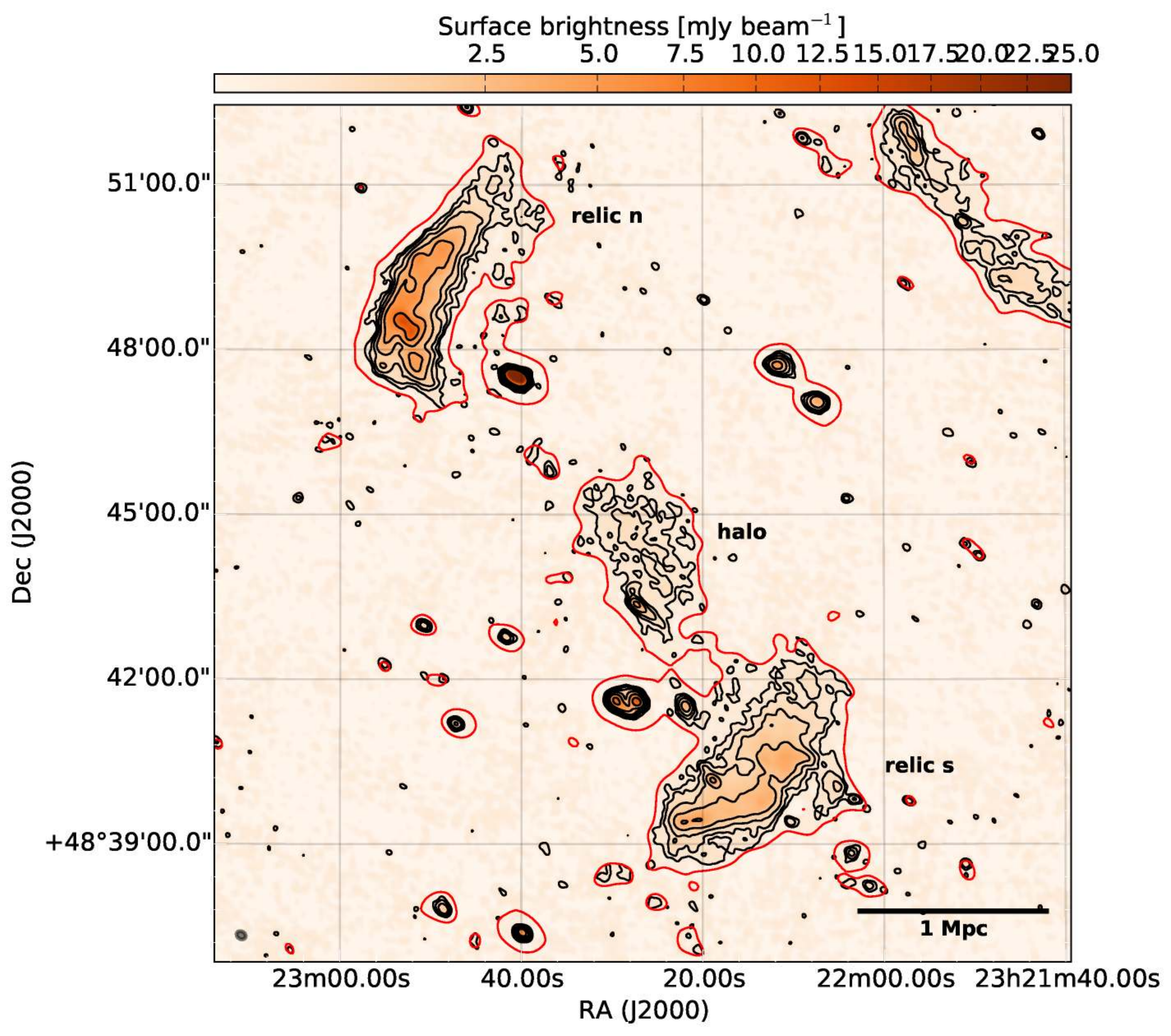}\\
\includegraphics[width=\columnwidth]{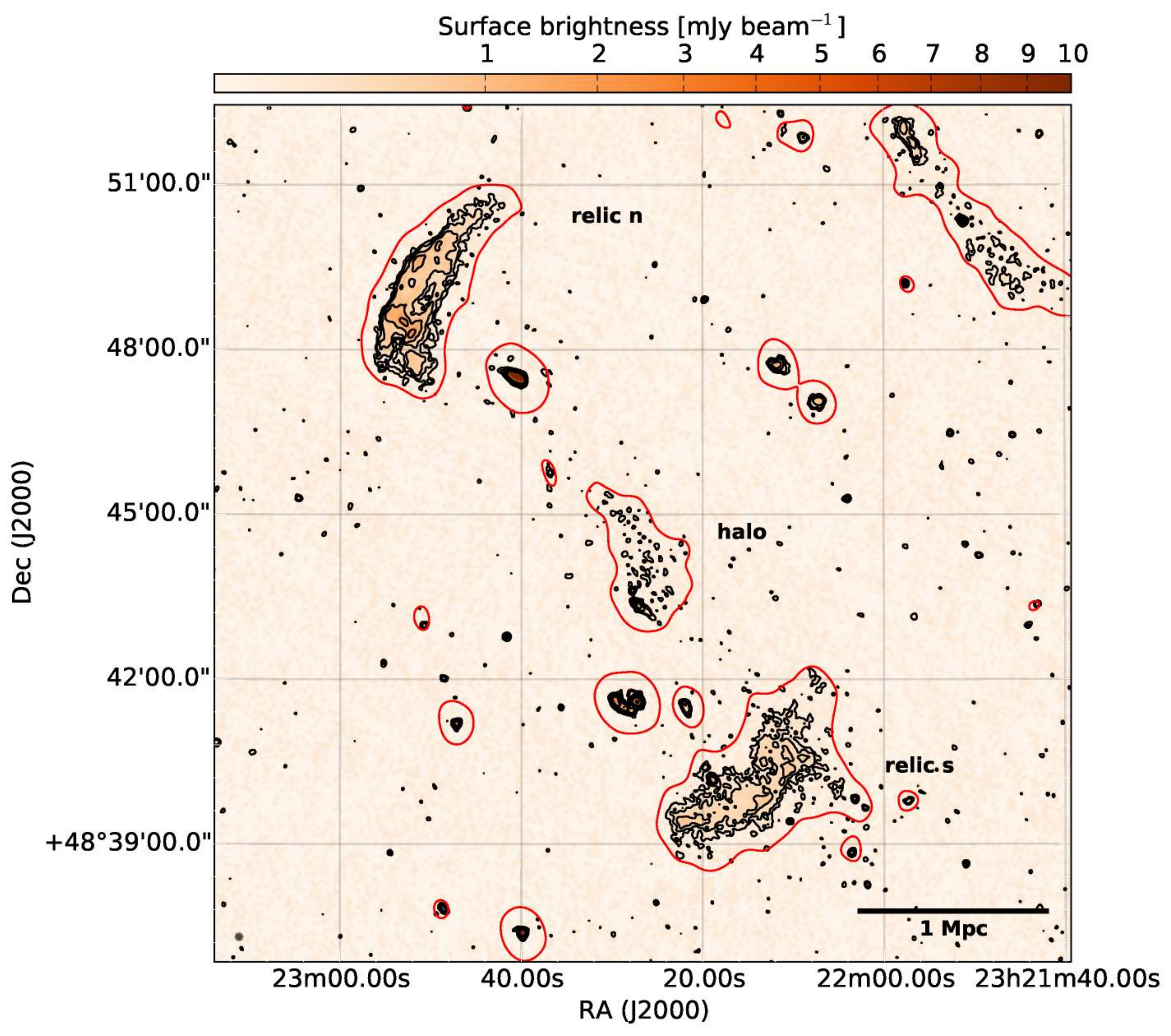}
\includegraphics[width=\columnwidth]{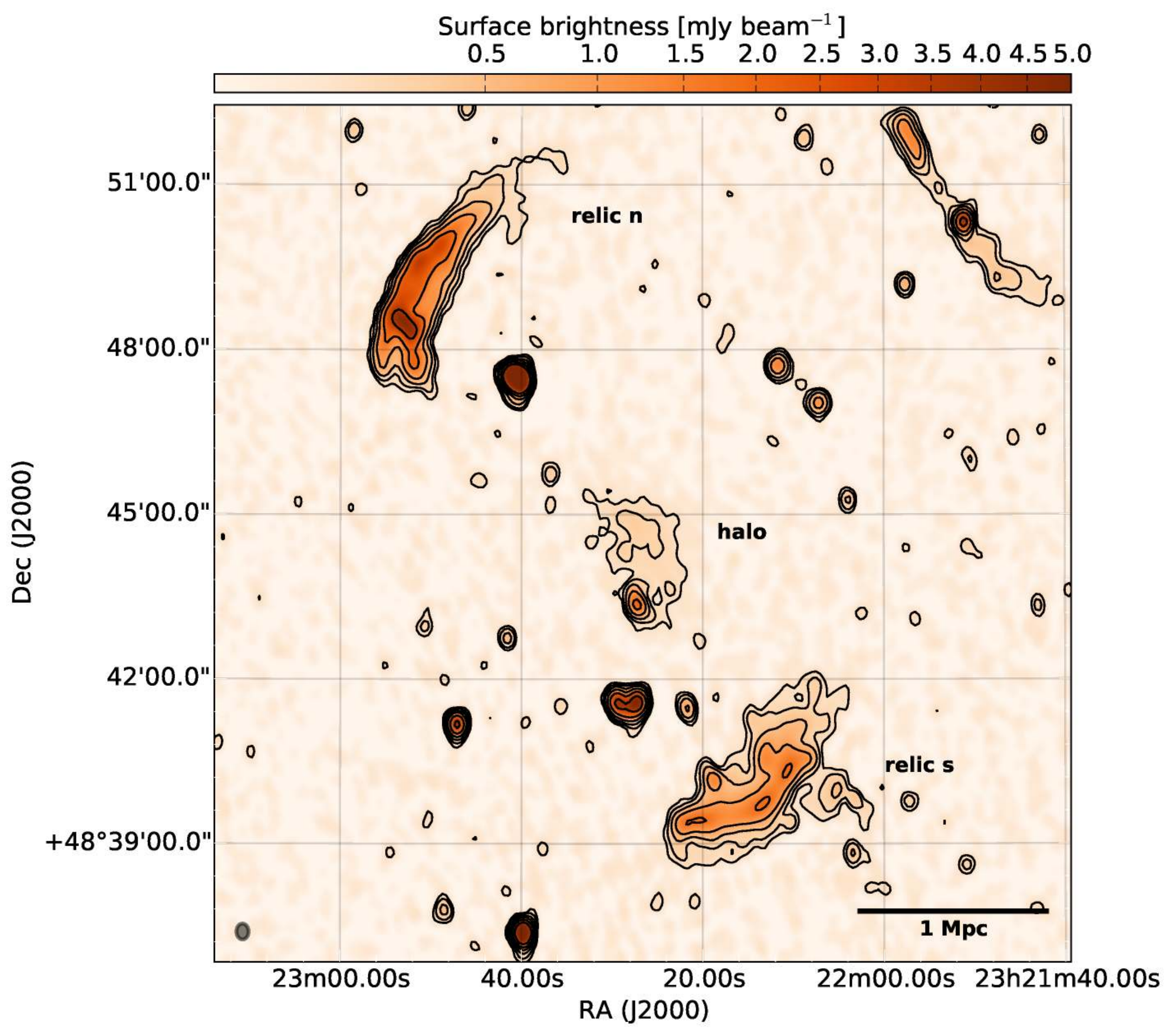}
\caption{Images of \target{} at multiple frequencies obtained with WSRT and GMRT. Red contours trace the low-resolution version; halo and relics are labelled. Top-left: GMRT image at 147 MHz, black contours at $\left(1,2,4,8,16,32\right)\times3 \sigma$ with $\sigma = 1.6$~\mjybeam (beam: \beam{27}{17}). Top-right: GMRT image at 323 MHz, black contours at $\left(1,2,4,8,16,32\right)\times3 \sigma$ with $\sigma = 86$~\mujybeam (beam: \beam{11}{8}); red contour at 600~\mujybeam ($3\sigma$ of the low-resolution map, beam: \beam{26}{20}). Bottom-left: GMRT image at 607 MHz, black contours at $\left(1,2,4,8,16,32\right)\times3 \sigma$ with $\sigma = 43$~\mujybeam (beam: \beam{6}{6}); red contour at 700~\mujybeam ($3\sigma$ of the low-resolution map, beam: \beam{30}{24}). Bottom-right: WSRT image at 1380 MHz, black contours at $\left(1,2,4,8,16,32\right)\times3 \sigma$ with $\sigma = 37$~\mujybeam (beam: \beam{17}{13}).}\label{fig:radio}
\end{figure*}

The data were calibrated with CASA\footnote{version 4.2, \url{http://casa.nrao.edu}} using the flux scale given by \cite{Perley2013}. The first step consisted of the removal of time ranges affected by shadowing. We then performed an initial bandpass correction using 3C147. Radio frequency interference was removed using the AOFlagger \citep{Offringa2012}. This initial bandpass correction prevents flagging data affected by the strong bandpass roll-off. After flagging, we again calibrated the bandpass and subsequently obtained complex gain solutions for the two calibrator sources. The channel-dependent polarization leakage terms were calibrated using the unpolarized source 3C137. The polarization angles were set using 3C286, carrying out a single correction per frequency channel. We then transferred all calibration solutions to the target source. For the target field, three rounds of phase-only self-calibration were performed. The data was imaged taking the spectral index into account during the deconvolution \cite[i.e., {\tt nterms=2}][]{Rau2011}. We used Briggs weighting with ${\rm robust}=0$ to gain sensitivity towards the diffuse emission. The final image was corrected for the primary beam attenuation. The final map is shown in Fig.~\ref{fig:radio} and has a resolution of \beam{17}{13} (rms: 37 \mujybeam). CASA is currently not able to perform a $T_{\rm sys}$ correction; this could cause a few percent offset in the absolute flux scale. We cross-checked the flux values of 11 point sources with fluxes $>10$~mJy visible in the NVSS survey \citep{Condon1998} finding our WSRT image systematically brighter by $\sim10\%$. We rescaled the final maps by this amount.  

\subsection{Radio: GMRT}
\label{sec:GMRT}

For each observing frequency, the visibility data was processed using an AIPS-based\footnote{AIPS: \url{http://www.aips.nrao.edu}} pipeline incorporating SPAM \citep[Source Peeling and Atmospheric Modelling][]{Intema2009} ionospheric calibration. The AIPS-based, semi-automated pipeline processing is very similar for each of the GMRT observing frequencies, being different only in the detailed settings of automated flagging. The pipeline uses the ParselTongue interface \citep{Kettenis2006} to access AIPS tasks, files and tables from Python. Flux and bandpass calibrations were derived from 3C~286 after three iterations of flagging and calibrating based on the \cite{Scaife2012} model. Additionally, instrumental phase calibrations were determined by filtering out ionospheric contributions \citep[see][]{Intema2009}, an important step for direction-dependent ionospheric calibration later on. Calibrations were transferred and applied to the target field data, simple clipping of spurious visibility amplitudes was applied, and data was averaged in time and frequency and converted to Stokes~{\textit I} to speed up processing. The effective bandwidths after flagging and averaging are 14.8, 31.3 and 29.2~MHz, centred on 147, 323 and 607~MHz, respectively. Self-calibration of the target field was started with an initial phase calibration using a multi-point source model derived from the NVSS, WENSS and VLSS radio source catalogues \citep{Condon1998, Rengelink1997, Cohen2007}, followed by (facet-based) wide-field imaging and CLEAN deconvolution of the primary beam area and several bright outlier sources (out to 5 primary beam radii). The visibility weighting scheme used during imaging is a mix between uniform and robust weighting (AIPS ROBUST -1), which generally provides a well-behaved point spread function (without broad wings) by down weighting the very dense central UV-coverage of the GMRT. Self-calibration was repeated three more times, including amplitude calibration in the final round, and outlier flagging on residual (source-model-subtracted) visibilities in between imaging and calibration. Next, two rounds of SPAM calibration and imaging were performed to minimize direction-dependent, residual phase calibration errors due to the ionosphere. In each round, (i) bright sources in and around the primary beam area were peeled \citep[e.g.,][]{Noordam2004}, (ii) the peeling phase solutions were fitted with a time-variable two-layer phase screen model, (iii) the model was used to generate ionospheric phase correction tables for the grid of viewing directions defined by the facet centres, and (iv) the target field was re-imaged and deconvolved while applying the appropriate correction table per facet. We refer to \citet{Intema2009} for more details. 

Given the large range of baseline lengths of the GMRT for the 607 and 323 MHz observations we produced two maps, one at high resolution ($\rm robust=0$) and one applying a Gaussian weight to the data (``tapering''). The final maps are shown in Fig.~\ref{fig:radio}. For the 147 MHz map we obtained a beam of \beam{27}{17} (rms: 1.6 \mjybeam). From the 323 MHz observation we extracted an high-resolution map (beam: \beam{11}{8}, rms: 86~\mujybeam) and a low-resolution map (\beam{26}{20}, rms: 200~\mujybeam) tapering at FWHM = 5 km. Finally, from the 607 MHz observation we obtained a high-resolution map (beam: \beam{6}{6}, rms: 43~\mujybeam) and a low-resolution map (\beam{30}{24}, rms: 230~\mujybeam) tapering at FWHM = 2 km.

\subsection{Optical: Keck}
\label{sec:opt}
We conducted a spectroscopic survey of the target cluster with the DEIMOS instrument on the Keck II 10\,m telescope on 2014, June 22. We used a similar instrument setup, slitmask design procedure, and reduction procedure to that detailed in \citet{Dawson2014}. Observations were taken using 1\arcsec\ wide slits with the 1200\,line\,mm$^{-1}$ grating, tilted to a central wavelength of 7400\,\AA, resulting in a pixel scale of $0.33$\,\AA\,pixel$^{-1}$, a resolution of $\sim1$\,\AA\ (50\,km\,s$^{-1}$), and typical wavelength coverage of 6000\,\AA\ to 8600\,\AA. The actual wavelength coverage may be shifted by $\sim\pm410$\AA\ depending where the slit is located along the width of the slitmask. For most cluster members this enabled us to observe H$\beta$, [\ion{O}{3}] 4960 \& 5008, \ion{Mg}{1} (b), \ion{Fe}{1},  \ion{Na}{1} (D), [\ion{O}{1}], H$\alpha$, and the [\ion{N}{2}] doublet. The position angle (PA) of each slit was chosen to lie between $\pm5\degr$ to 30$\degr$ of the slitmask PA to achieve optimal sky subtraction\footnote{\url{http://astro.berkeley.edu/~cooper/deep/spec2d/slitmask.html}} during reduction with the DEEP2 version of the spec2d package \citep{Newman2013}. Within this range the slit PA was chosen to minimize the effects of chromatic dispersion by the atmosphere by aligning  the slit, as much as possible, with the axis connecting the horizon, object and zenith \citep[see e.g.][]{Filippenko1982}. We observed a total of two slit masks with approximately 120 slits per mask. For each mask we took three 900\,s exposures.

Our primary objective for the spectroscopic survey was to maximize the number of spectroscopic redshifts of cluster members. This was challenging since the only available optical imaging data was from the Digitized Sky Survey\footnote{\url{https://archive.stsci.edu/cgi-bin/dss_form}} (DSS), which has a limiting magnitude of $\sim22$ and $\sim3.5\arcsec$ full-width-half-max point spread function. Since the archival DSS catalogues do not push faint enough to detect a significant number of the cluster galaxies we ran Sextractor \citep{Bertin1996} on the DSS blue, red, and infrared digitized images following a similar setup and procedure to that detailed in \citet{Jee2015}. We found that the poor seeing of the DSS imaging coupled with the low galactic latitude ($b=-11$\,deg) made star galaxy separation based on object shape impractical and selected spectroscopic targets based purely on color. We visually inspected the DSS color composite images to determine that \textit{blue -- red} $>-2$ appeared to exclude the majority of blue stars while being inclusive of apparent cluster galaxies. The slitmask creation procedure and reduction of the raw data are detailed in \citet{Dawson2014}.

We obtained 291 secure redshifts, however only 60 are of galaxies with the remaining 231 of stars; indicative of the importance of deep subarcsec seeing imaging for spectroscopic target selection in low galactic fields. Of the galaxy spectra 42 are likely cluster members, defined as within 3000 km s$^{-1}$ (three times the velocity dispersion) of the mean cluster redshift $z_{\rm cluster}=0.335$. Of the remaining spectroscopic galaxies 13 are foreground and 5 are background.

\section{PSZ1 G108.18-11.53}
\label{sec:target}

\target{} was initially mentioned in \cite{PlanckCollaboration2013} but no redshift was known. In \cite{PlanckCollaboration2014} the cluster redshift was measured to be $z=0.336$. Independently, with Keck observations, we measured a spectroscopic redshift of $0.33466^{+0.00061}_{-0.00061}$ (Sec.~\ref{sec:gal_distrib}). Throughout this paper we will use our redshift value.

\begin{figure}
\centering
\includegraphics[width=\columnwidth]{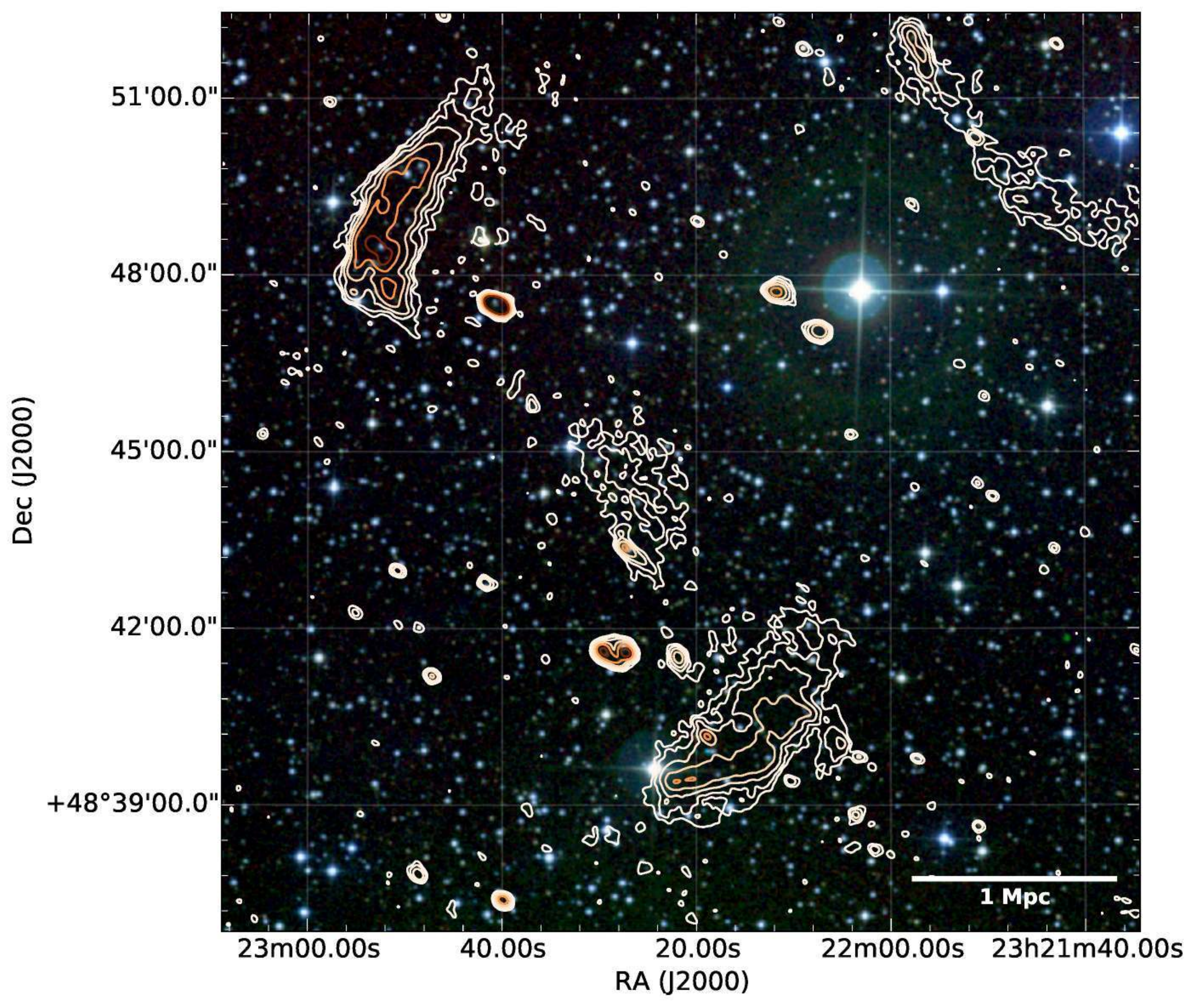}
\caption{Composite image using the infrared, red and blue band of the DSS. Superimposed contours from the 323 MHz image obtained with the GMRT (contour levels as in Fig.~\ref{fig:radio}, radio emission at 323 MHz).}\label{fig:dss}
\end{figure}

The cluster has been initially discovered by measuring its SZ signal over the cosmic microwave background radiation. The SZ signal integrated over the clusters angular extent measures the total thermal energy of the cluster gas, which is tightly related to cluster mass \citep[e.g.][]{Borgani2006}. For \target{} this has been measured to be $M_{500} = (7.7\pm0.6) \times 10^{14}$~\Msun \citep{PlanckCollaboration2015}. At these masses the Planck SZ-catalogue is complete at $>80\%$ and only 72 more massive clusters are present in the catalogue.

\begin{figure}
\centering
\includegraphics[width=\columnwidth]{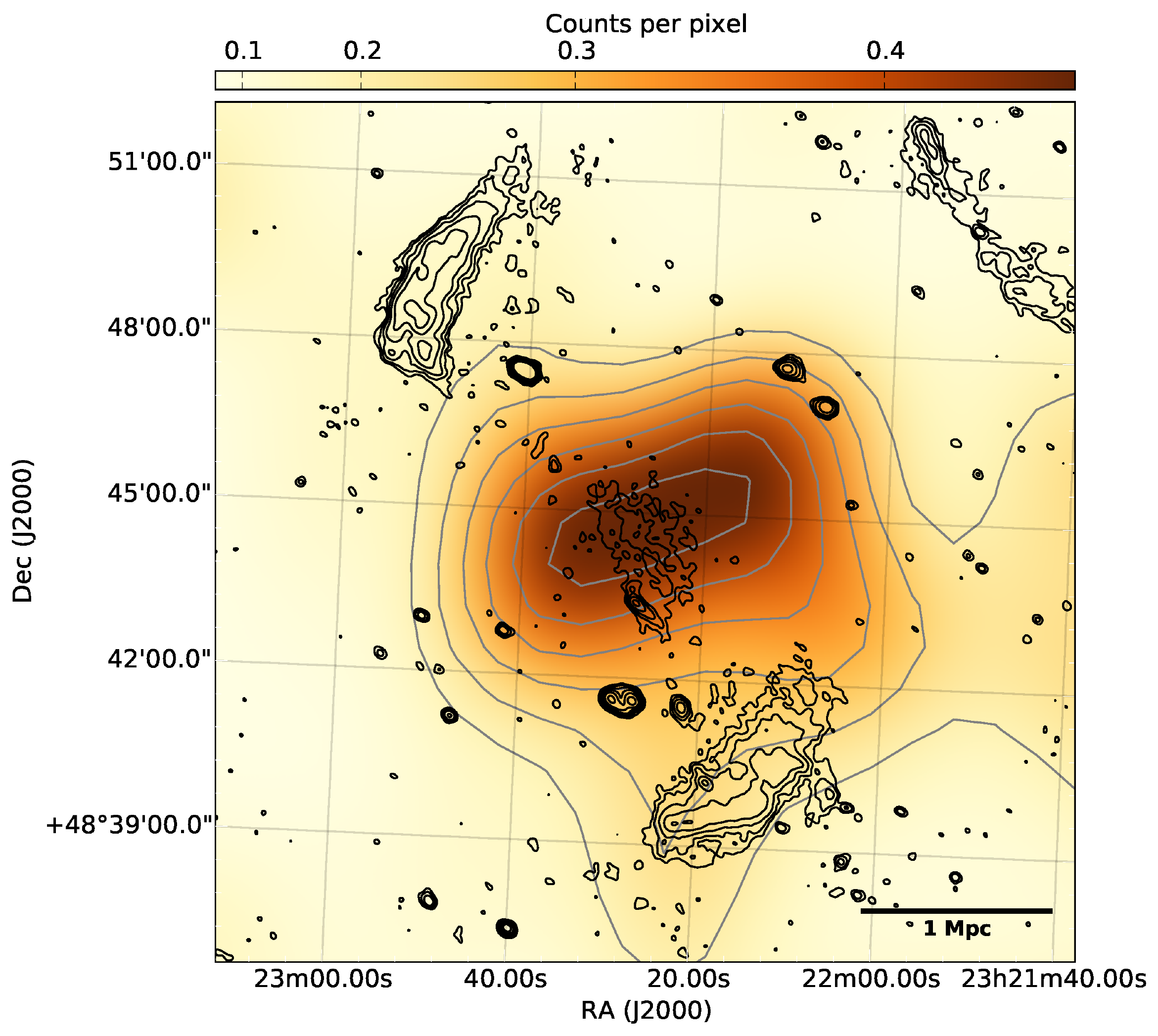}
\caption{Broad band ($0.1-2.4$ keV) photon image from the ROSAT all sky survey (contour levels as in Fig.~\ref{fig:radio}, radio emission at 323 MHz). Image has been smoothed with a two-dimensional Gaussian kernel of $\sigma=2$~pixels (1~$\textrm{pixel} = 45$~arcsec). Total estimated X-ray luminosity is $L_{\rm X} \sim 7.5 \times 10^{44}$ erg s$^{-1}$.}\label{fig:rosat}
\end{figure}

\target{} is also detected in ROSAT (see Fig.~\ref{fig:rosat}), which provide an approximate estimation of the total cluster luminosity: $L_{\rm X} \sim 7.5 \times 10^{44}$ erg s$^{-1}$. Unfortunately, the angular resolution of ROSAT prevents us from carrying out a further analysis of morphological features. Using the ($L$, $T$) relations provided by \cite{Pratt2009} we estimated a cluster temperature of T$\sim6.5$~keV.

\begin{table}
\centering
\begin{threeparttable}
\begin{tabular}{ll}
\multicolumn{2}{c}{PSZ1 G108.18-11.53}\\
\hline
Redshift & $0.33466^{+0.00061}_{-0.00061}$ \bigstrut[t]\\
Coordinates & RA: \hms{23}{22}{29.7} -- Dec: \dms{+48}{46}{30} \\
X-ray luminosity\tnote{a} & $7.5 \times 10^{44}$ erg s$^{-1}$ \\
$M_{500}$\tnote{a} & $ (7.7\pm0.6) \times 10^{14}$ \Msun\\
Velocity dispersion & $910^{+120}_{-90}\,\mathrm{km}\,\mathrm{s}^{-1}$\\
\hline
\end{tabular}
\begin{tablenotes}
    \item[a] $0.1 - 2.4$ keV, ROSAT
    \item[b] From SZ measurement, \cite{PlanckCollaboration2015}
\end{tablenotes}
\end{threeparttable}
\caption{Cluster properties}\label{tab:cluster}
\end{table}

\subsection{Galaxy distribution}
\label{sec:gal_distrib}
  
\begin{figure}
\centering
\includegraphics[width=.8\columnwidth]{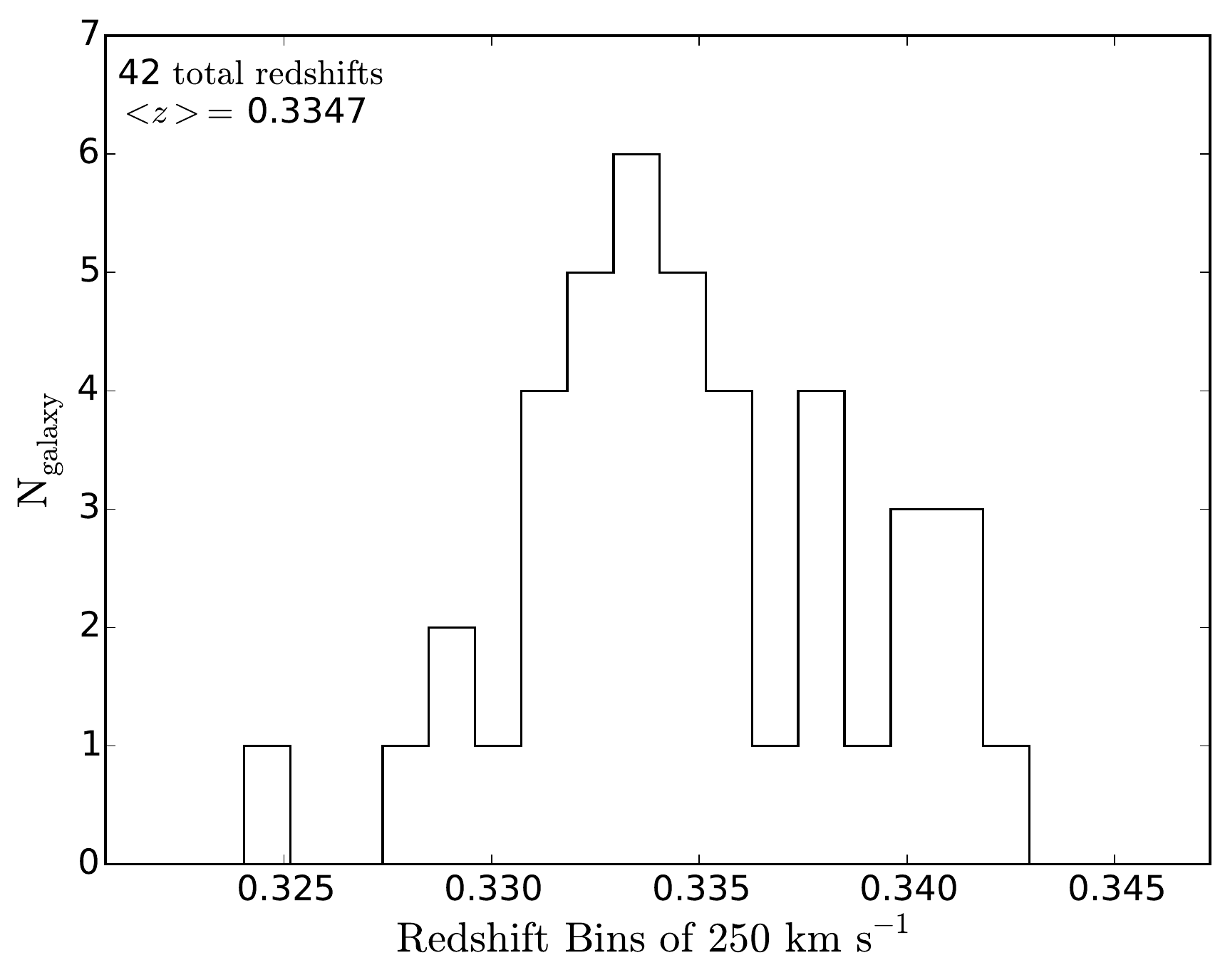}
\caption{Distribution of spectroscopic measurements. Average value is at $0.33466^{+0.00061}_{-0.00061}$.}\label{fig:z-distrib}
\end{figure}

\begin{figure*}
\centering
\includegraphics[width=\columnwidth]{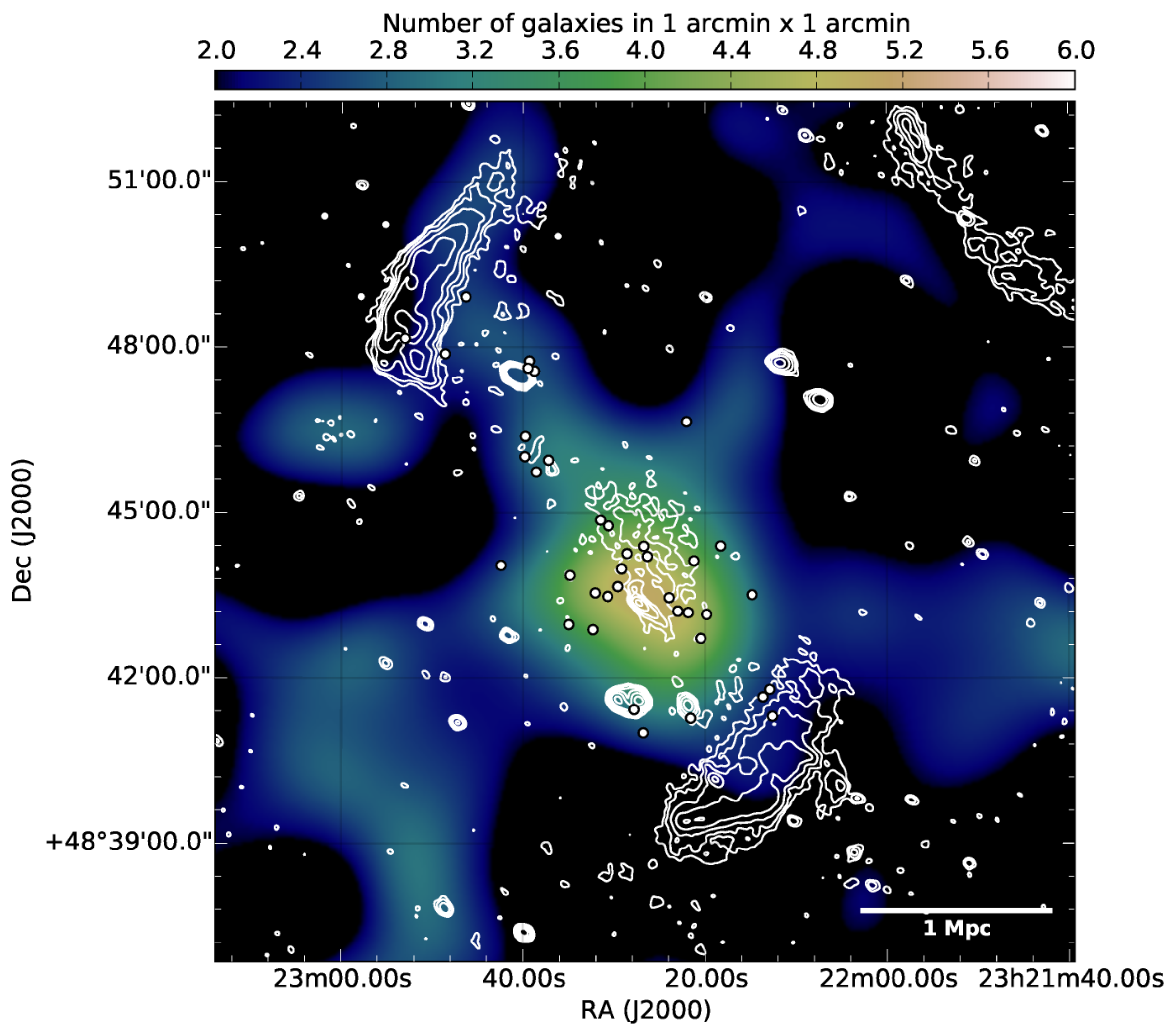}
\includegraphics[width=\columnwidth]{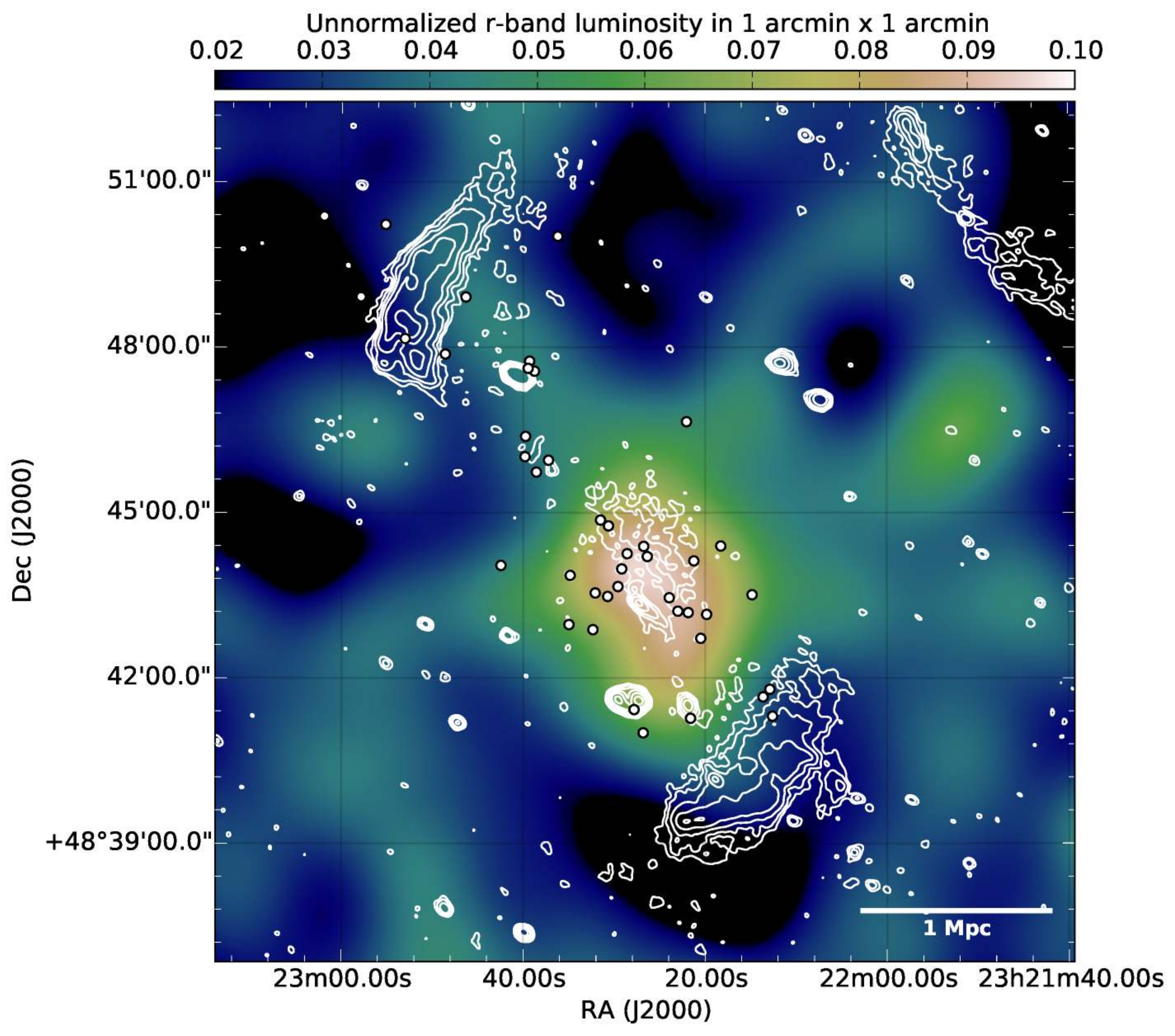}
\caption{Cluster galaxy number density distribution (left) and unnormalized luminosity density distribution (right) in the field of \target{} (contour levels as in Fig.~\ref{fig:radio}, radio emission at 323 MHz). Maps are based on DSS galaxies with a cluster membership selection trained on the spectroscopically confirmed cluster members. The map has been smoothed by a Gaussian with $1\sigma$ radius = 50\arcsec. White points are the cluster galaxies with spectroscopic data.}\label{fig:gal_distrib}
\end{figure*}

To investigate the distribution of the cluster galaxies we consider all 42 galaxy spectra within the redshift range $0.3217 < z < 0.3483$, which is  $z_\mathrm{cluster}\pm 3\times\sigma$, where $z_\mathrm{cluster}=0.335$ and $\sigma$ is the approximate velocity dispersion ($1000\,\mathrm{km}\,\mathrm{s}^{-1}$). We show the redshift distribution of these galaxies in Fig.~\ref{fig:z-distrib}. We estimate the cluster redshift and velocity dispersion using the biweight-statistic and bias-corrected 68\% confidence limit \citep{Beers1990} applied to 100,000 bootstrap samples of the cluster spectroscopic redshifts. We find a cluster redshift of $0.33466^{+0.00061}_{-0.00061}$ and a velocity dispersion of $910^{+120}_{-90}\,\mathrm{km}\,\mathrm{s}^{-1}$. We caution the reader that care should be taken when interpreting these results given the small number of spectroscopic redshifts and that we have treated a likely major merger of two or more subclusters as a single system.
We perform a Kolmogorov--Smirnov test to compare the redshift distribution relative to the normal distribution defined by the calculated bi-weight location (redshift) and scale (velocity dispersion) and find a p-value  of 0.92. Thus the redshift distribution is consistent with being a Normal, providing no evidence of significant substructure along the line of sight.

To examine the distribution of the cluster galaxies in projection we use a color-magnitude selection based on the DSS photometry. We do not use the DEIMOS spectroscopic galaxies as there is a strong selection bias due to the limited number of spectroscopic masks observed and their limited field of view ($5\arcmin \times 16\arcmin$). Instead we use the spectroscopic galaxies to inform a reasonable color-magnitude selection. Considering just the spectroscopic objects this selection results in 34 cluster galaxies, 5 foreground galaxies, 3 background galaxies, and 67 stars. Thus there would be a large (Poisson) background component in any resulting galaxy density map, reflective of the poor imaging data. We show the photometric galaxy number density map in Fig.~\ref{fig:gal_distrib}. This map shows a general elongated over-density along the axis connecting the two radio relics, with the peak of the over-density located near the radio halo and the peak of the X-ray emission. This distribution is compatible with the idea of two sub-clusters which merged along a merger axis perpendicular to the line of sight and oriented along a line connecting the two radio relics. It is also interesting that the galaxy luminosity density map shows a smaller over-density in the west, which may be related to the east-west elongated X-ray mission (Fig.~\ref{fig:rosat}), although given the large background in the galaxy density map this is speculative.

\subsection{Radio relics}
\label{sec:relics}

Two large radio relics are visible on the north-east and south-west side of the cluster. Their distance from the peak of the ROSAT X-ray emission is of $\sim1750$ kpc and $\sim1280$ kpc respectively. Their relative distance is of $\sim 3100$ kpc. The luminosity of the northern relic at 1400 MHz ($k$-corrected) is $(26.6\pm0.1) \times 10^{24}$ W/Hz, while the southern relic luminosity is $(18.0\pm0.1) \times 10^{24}$ W/Hz\footnote{Flux errors for extended sources are computed as $S_{\rm err} = \sigma \times \sqrt{N_{\rm beam}}$, where $\sigma$ is the local image rms and $N_{\rm beam}$ is the number of beams covering the source extension.}. The northern relic is 33\% more luminous than the southern although their extension is similar (1500 and 1300 kpc, respectively). The northern relic brightness is not uniform across its extension peaking towards the south-east side. The external edge is slightly curved (a common feature for radio relics) while the internal edge is straighter. Both internal and external edges show quite steep brightness gradients. The relic is thicker towards the south (390 kpc) and it becomes narrower towards north. Conversely, the southern relic brightness is uniform along its extension, but the source looks less compact and with less defined edges. The external edge has a quite steep brightness gradient, although half way along the relic extension, a filament of 240 kpc is visible in front of the relic. The relic thickness is uniform and of about 490 kpc. The surface brightness of the inner edge of the relic mildly decreases moving towards the cluster centre and its full extension is likely not entirely visible.

\begin{table*}
\centering
\begin{threeparttable}
\begin{tabular}{llllllll}
Name    & Flux 147 MH\tnote{a}z & Flux 323 MHz\tnote{a} & Flux 607 MHz\tnote{a} & Flux 1380 MHz\tnote{a} & Spectral       & Luminosity\tnote{b} & LLS \\
        & (mJy)        & (mJy)        & (mJy)        & (mJy)         & index          & ($10^{24}$ W/Hz) & (kpc) \\
\hline
Relic N & $1032\pm14$  & $422\pm2$    & $177\pm2$    & $67.7\pm0.3$  & $-1.25\pm0.02$ & $26.6\pm0.1$ & 1500 \bigstrut[t]\\
Relic S & $753\pm15$   & $323\pm3$    & $128\pm2$    & $45.4\pm0.3$  & $-1.28\pm0.02$ & $18.0\pm0.1$ & 1300 \\
Halo    & $124\pm11$    & $56\pm2$     & $20\pm1$     & $6.8\pm0.2$   & $-1.40\pm0.07$ & $2.8\pm0.1$  & 850 \\
\end{tabular}
\begin{tablenotes}
    \item[a] fluxes are all extracted from a region equal to the $3\sigma$ level of the 323 MHz map.
    \item[b] at 1400 MHz, k-corrected and spectral-index rescaled.
\end{tablenotes}
\end{threeparttable}
\caption{Radio properties}\label{tab:relics}
\end{table*}

\begin{figure}
\centering
\includegraphics[width=\columnwidth]{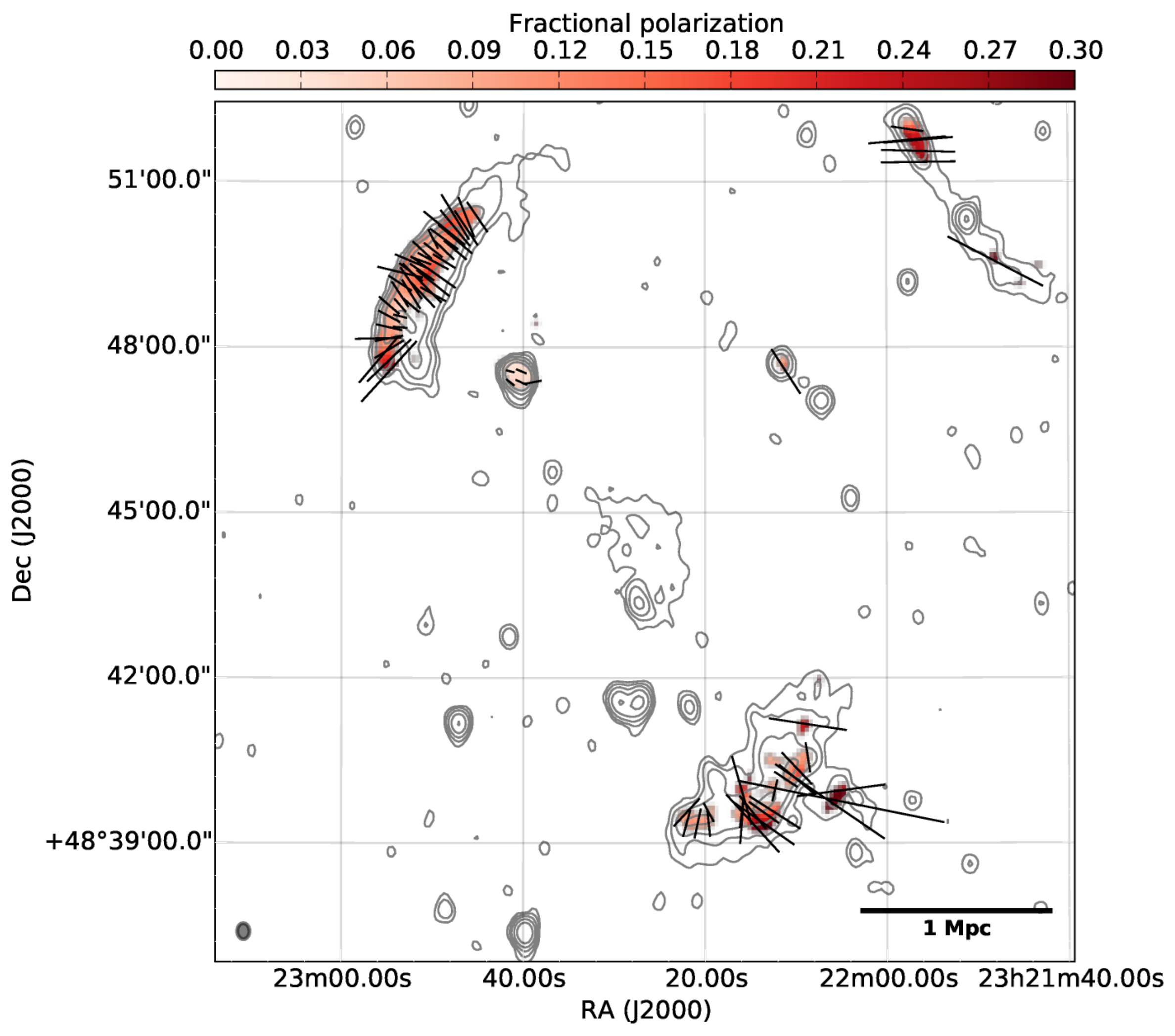}
\caption{Fractional polarization map with E-vectors displayed. The polarization angle has been corrected for galactic Faraday rotation. Contour levels as in Fig.~\ref{fig:radio}, radio emission at 1380 MHz.}\label{fig:WSRT_F}
\end{figure}

A $10-30$\% polarization fraction is detected in the brightest regions both in the northern and southern relics (see Fig.~\ref{fig:WSRT_F}). Using the map provided by \cite{Oppermann2012} we estimated a Faraday depth in the direction of \target{} of $-6.5$ rad m$^{-2}$. The polarization angle has therefore been corrected for galactic Faraday rotation by adding an angle of 17.5\deg. The northern relic magnetic field appears aligned to the relic extension and curves close to the southern end. The southern relic polarization is less ordered, but it shows again a magnetic field generally aligned with the relic extension. This magnetic field orientation might arise as a consequence of field compression by the shock wave which formed the radio relic \citep[see e.g.][]{Iapichino2012}, it could come from magnetic field amplification from CRs turbulence \citep{Bruggen2013}, or it may be a prerequisite for efficient particle acceleration \citep{Guo2014}.

\subsection{Radio halo}
\label{sec:halo}

\target{} has a centrally located radio halo. The source is quite uniform in brightness, and slightly elongated along the cluster merging axis. The halo maximum extension is $\sim850$ kpc, with an average diameter of $\sim 675$ kpc\footnote{Measured following \cite{Cassano2007} as $\sqrt{d_{\rm min} \times d_{\rm max}}$, $d_{\rm min}$ and $d_{\rm max}$ being the minimum and maximum diameter measured on the $3\sigma$ radio isophotes of the 323 MHz map.}, and a luminosity of $(2.8\pm0.1) \times 10^{24}$~W/Hz at 1400 MHz ($k$-corrected). The halo has a rather steep spectral index of $\alpha=-1.40\pm0.07$, close to the definition of ultra-steep spectrum radio halo \citep{Cassano2013}. The halo lies slightly below the cluster mass--power correlation found by \cite{Cassano2013} as already noticed for other halos with a very steep spectrum. Given its luminosity, the radio halo is slightly smaller than expected from scaling-relations. \cite{Cassano2007} measured halo size and luminosity for 15 radio halos finding a clear trend for more luminous radio halo to be more extended. For the luminosity of the halo in \target{}, the expected radio halo average diameter is $\sim 1000$ kpc. Polarization is not detected down to the level of 10\%.


\subsection{Radio environment}

\begin{figure*}
\centering
\includegraphics[width=\textwidth]{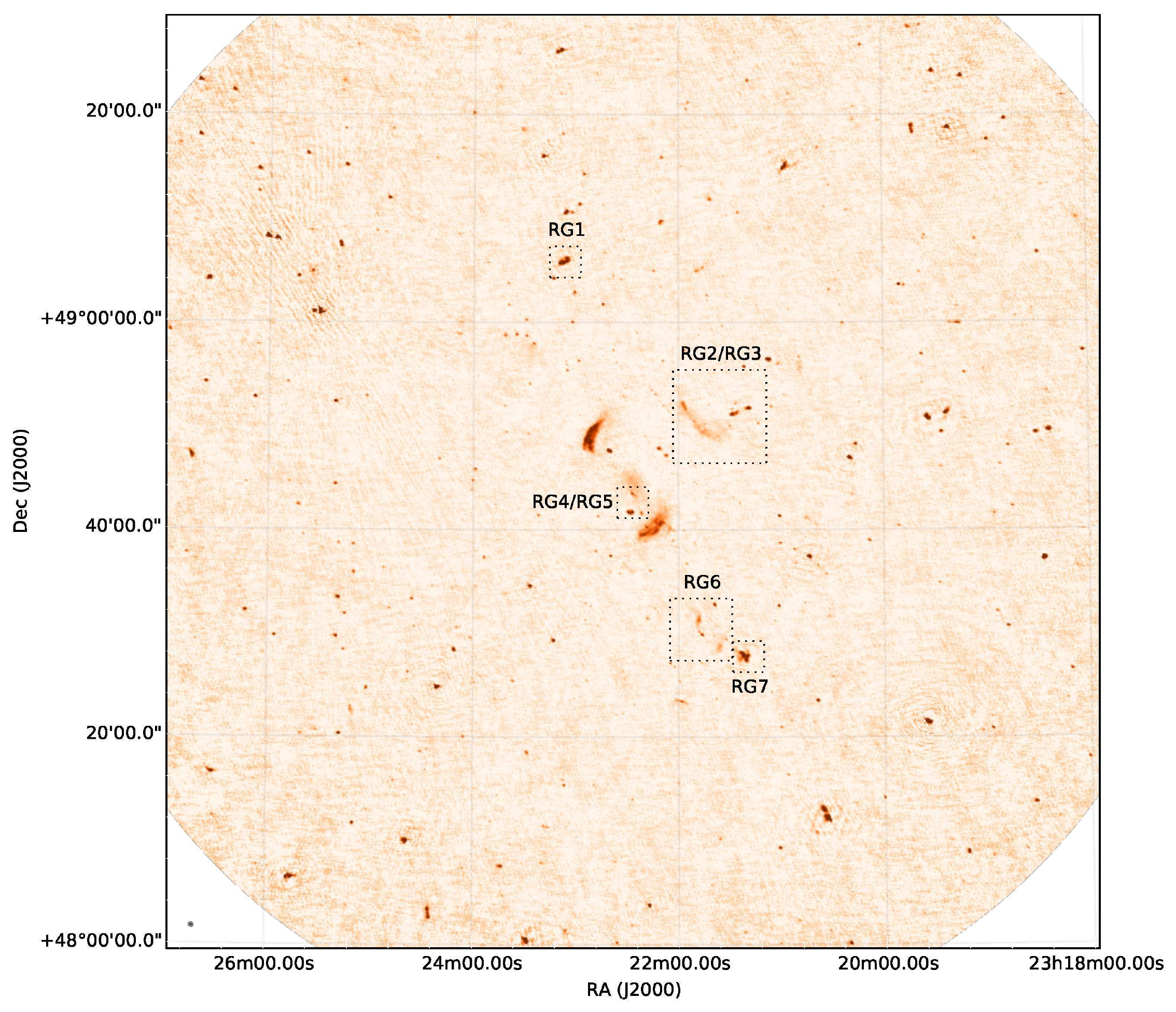}
\caption{Entire field of view at 323 MHz as seen by the GMRT, highlighted the regions zoomed in Fig.~\ref{fig:zoom}. Image noise at the field centre is 100~\mujybeam, resolution is \beam{10}{8}.}\label{fig:GMRT_330-large}
\end{figure*}

\begin{figure*}
\centering
\includegraphics[width=.45\textwidth]{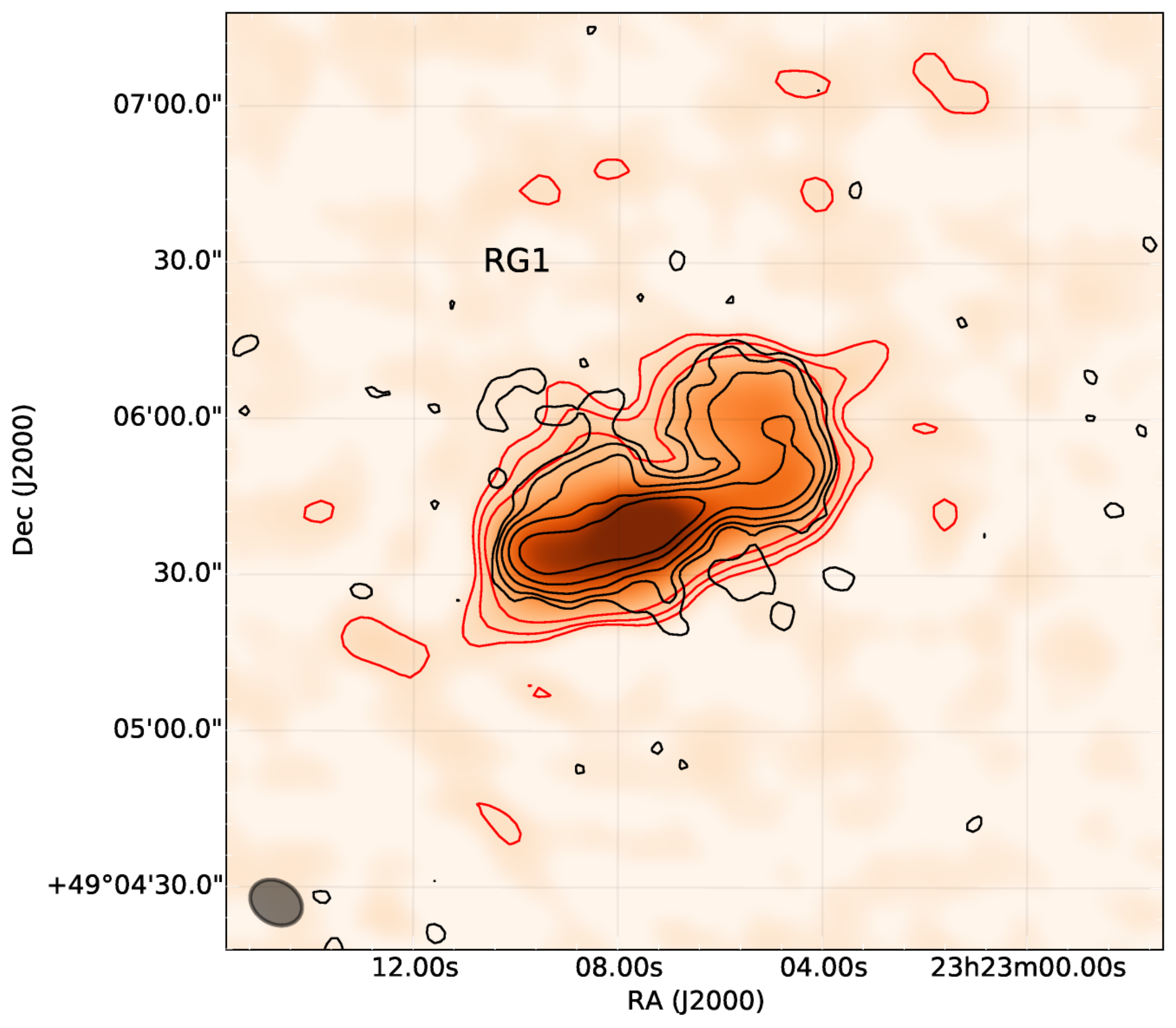}
\includegraphics[width=.45\textwidth]{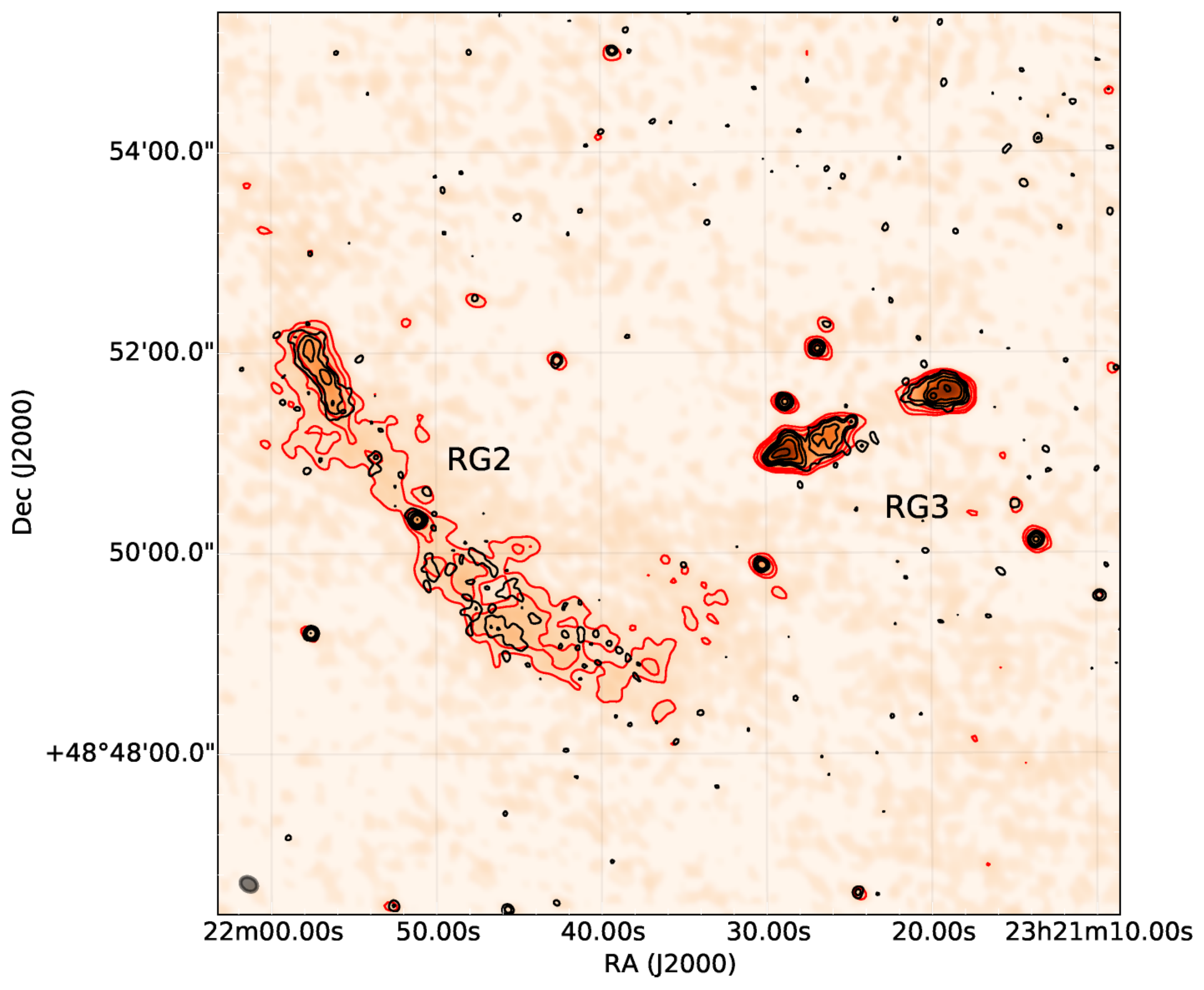}\\
\includegraphics[width=.45\textwidth]{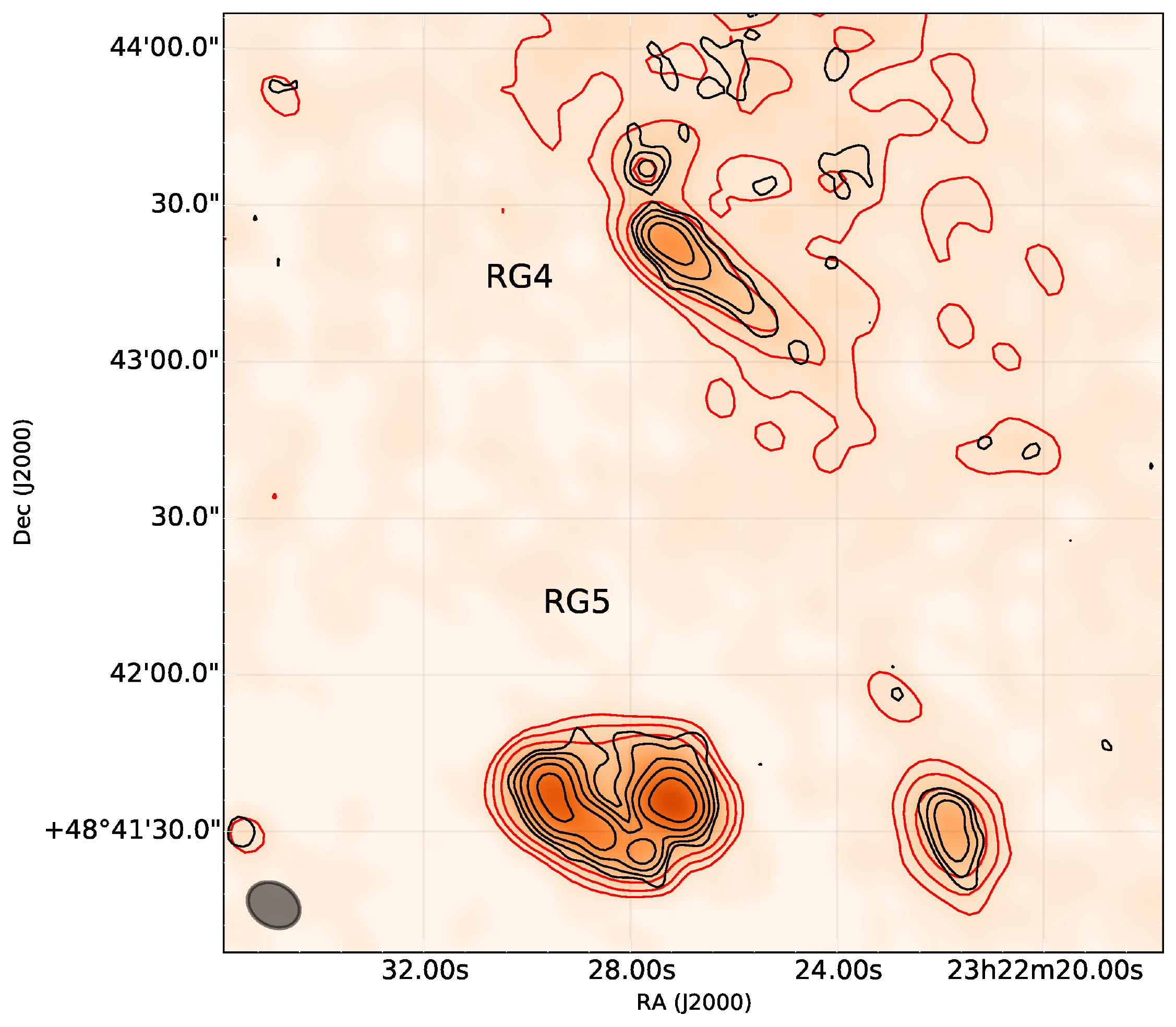}
\includegraphics[width=.45\textwidth]{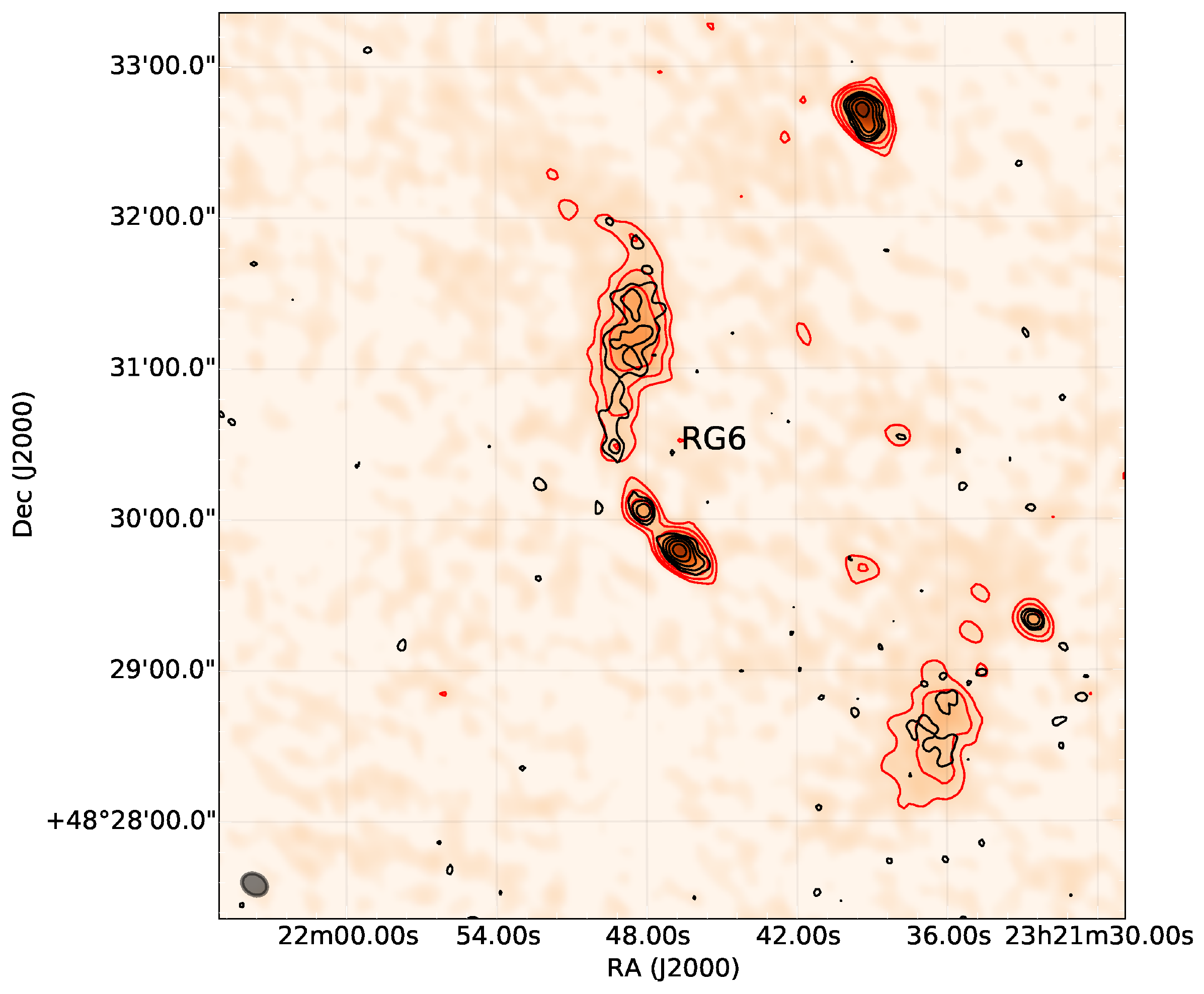}\\
\includegraphics[width=.45\textwidth]{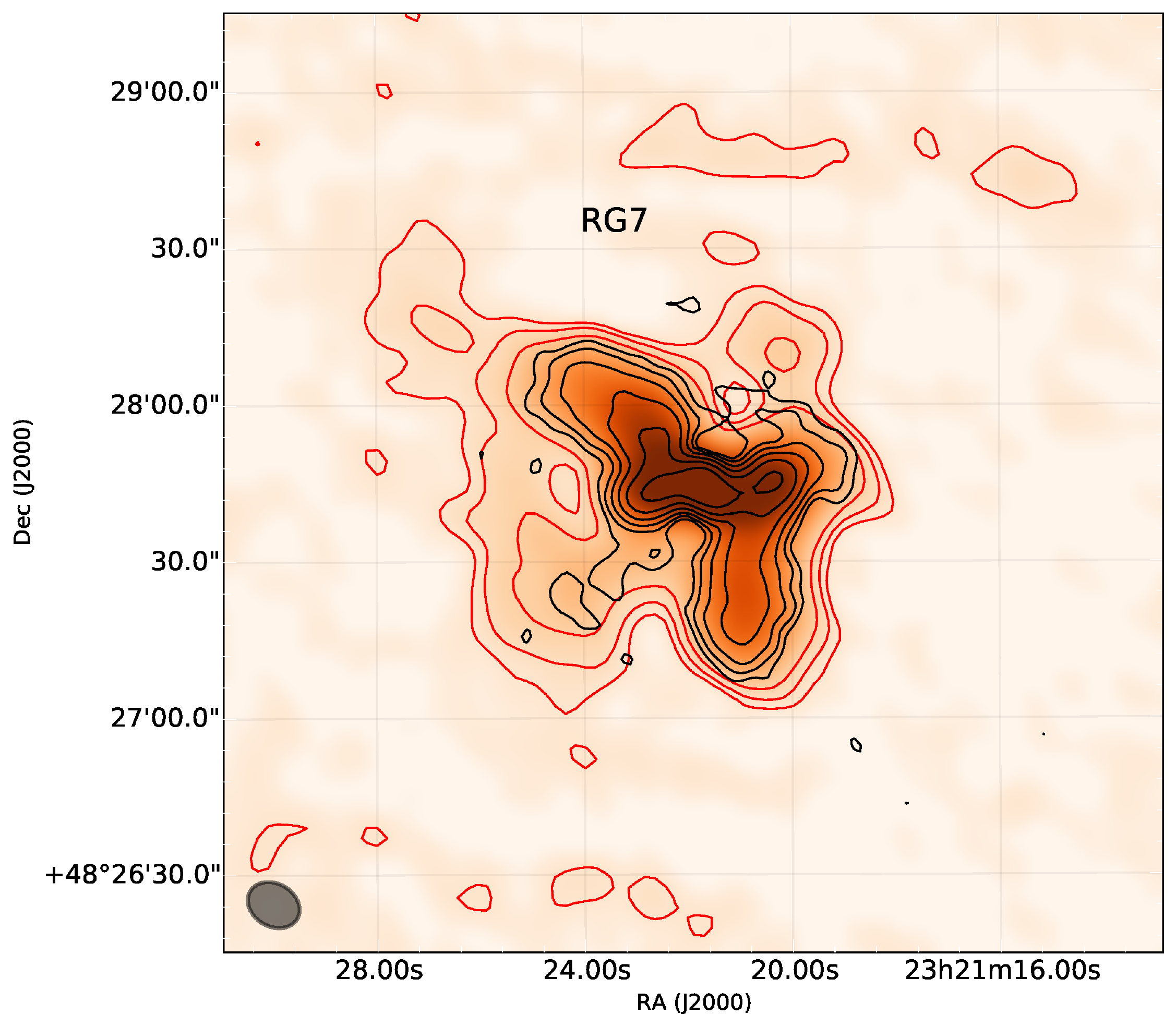}
\caption{Zoomed images on some resolved sources in the 323 MHz field as observed with the GMRT. Given the different location across the map the rms values vary. Black contours trace the 607 MHz emission at $\left(1,2,4,8,16,32\right)\times3 \sigma$ (from top-left $\sigma = 62, 47, 47, 62, 87$~\mujybeam, resolution \beam{5}{5}), red contours show the emission at 323 MHz at $\left(1,2,4\right)\times3 \sigma$ (from top-left $\sigma = 130, 85, 100, 100, 100$~\mujybeam, resolution \beam{10}{8}).}\label{fig:zoom}
\end{figure*}

\begin{figure}
\centering
\includegraphics[width=\columnwidth]{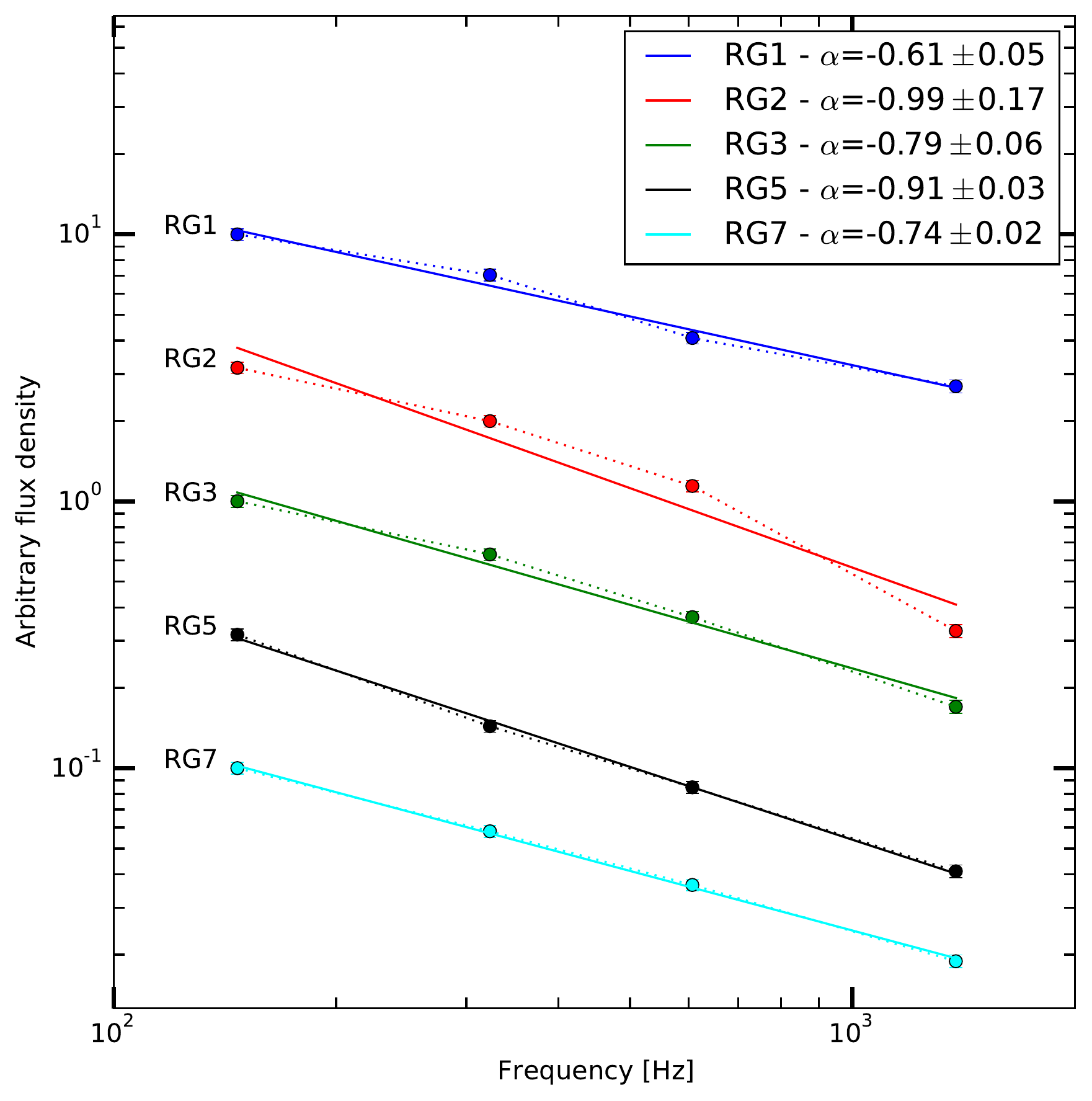}
\caption{Global spectral index analysis for the radio galaxies listed in Table~\ref{tab:RG} (RG4 and RG6 are excluded). Solid lines are linear regressions, dotted lines connect data points. Fluxes have been arbitrarily rescaled to show all the spectra in a single plot retaining their shape in the log-log space. Error bars are at 5\% of the data point value.}\label{fig:globalspidxRG}
\end{figure}

The radio environment in the region of \target{} is particularly rich with interesting sources. In Fig.~\ref{fig:GMRT_330-large} we show the full field of view of the GMRT 323 MHz observation. Apart from the two radio relics and the radio halo a number of extended radio galaxy (RG) are also present. The zoomed high and low-resolution images of these radio galaxies are in Fig.~\ref{fig:zoom}, while their properties are summarized in Table~\ref{tab:RG}. Global spectral indices calculated using all four frequencies (147, 323, 607, 1380 MHz) are shown in Fig.~\ref{fig:globalspidxRG}.

\begin{table*}
\centering
\begin{threeparttable}
\begin{tabular}{llllllll}
Name & RA       & DEC      & Flux 147 MHz & Flux 323 MHz & Flux 607 MHz & Flux 1380 MHz & Spectral\\
     & hh:mm:ss & dd:mm:ss & (mJy)        & (mJy)        & (mJy)        & (mJy)         & index   \\
RG1  & 23:23:08 & 49:05:39 & 247          & 174          & 101          & 67            &  $-0.61\pm0.05$ \bigstrut[t]\\
RG2  & 23:21:51 & 48:50:23 & 122          & 77           & 44           & 13            &  $-0.99\pm0.17$ \\
RG3  & 23:21:24 & 48:51:16 & 106          & 67           & 39           & 18            &  $-0.79\pm0.06$ \\
RG4  & 23:22:27 & 48:43:22 & --           & 12           & 6.5          & --            &          \\
RG5  & 23:22:28 & 48:41:25 & 97           & 44           & 26           & 13            &  $-0.91\pm0.03$ \\
RG6 (core/jet) & 23:21:46 & 48:29:48 & $<16$\tnote{a}& 13& 9.8          & 7.0           &          \\
RG6 (lobe N) & 23:21:48 & 48:31:11 & --   & 23           & 17           & 6.5           &          \\
RG6 (lobe S) & 23:21:36 & 48:28:32 & --   & 12           & $>7.4$\tnote{b} & 1.5        &          \\
RG7  & 23:21:22 & 48:27:43 & 400          & 232          & 146          & 76            &  $-0.74\pm0.02$ \\
\end{tabular}
\begin{tablenotes}
    \item[a] Source is blended with its northern lobe.
    \item[b] Source is only partially detected.
\end{tablenotes}
\end{threeparttable}
\caption{Radio galaxies}\label{tab:RG}
\end{table*}

\paragraph*{RG1}
Searching the NED archive we found a possible optical counterpart for RG1: 2MASX J23230751+4905386. Unfortunately the redshift of this galaxy is unknown but it is likely too far apart to be a cluster member and it is probably a foreground source. The global spectral index of this RG is $\alpha = -0.58\pm0.04$ and its emission is likely dominated by the central core. The extended lobes can be classified as an FR I with an interesting abrupt bending of one of the lobes similar to those of Z-shaped FR II radio galaxies \citep{Gopal-Krishna2003}.
\paragraph*{RG2}
No optical counterpart has been found for this RG but given its extension and location it is likely a foreground object. This source is a rare example of a hybrid FR I/II radio galaxies with one FR I lobe and one FR II lobe \citep[hybrid morphology radio sources, ``HyMoRS''][]{Gopal-Krishna2000}. In depth analysis of this source will be presented in a forthcoming paper.
\paragraph*{RG3}
No optical counterpart has been found for RG3 which is an FR II radio galaxy with a global spectral index of $\alpha=0.77\pm0.09$. The spectral index (see Fig.~\ref{fig:globalspidxRG}) shows signs of ageing as it becomes steeper at higher frequencies.
\paragraph*{RG4}
Although no optical counterpart was found for RG4, its location and clear interaction with the ICM makes it a probable member of the cluster. The source is an head-tail radio galaxy whose lobes are bent by their interaction with the surrounding medium.
\paragraph*{RG5}
RG5 has an identified optical counterpart located at RA:350.61\deg, DEC:48.69\deg ($z=0.333$). The source is classified as a cluster member following the criteria of Sec.~\ref{sec:opt}. The source is a wide angle tail radio galaxy, also common in the proximity of galaxy clusters.
\paragraph*{RG6}
No optical counterpart has been identified for RG6. The source might be another example of HyMoRS although the second lobe does not show a clear hotspot presence. In depth analysis of this source will be presented in a forthcoming paper.
\paragraph*{RG7}
RG7 has an identified optical counterpart in 2MASX J23212159+4827436. No redshift is available. The source appears to be a Z-shaped radio galaxy with prominent wings. Interestingly on the larger scales emission is also detected on the other side of the lobes compared to the wings direction.

\section{Spectral analysis}
\label{sec:spanalysis}

\begin{figure}
\centering
\includegraphics[width=\columnwidth]{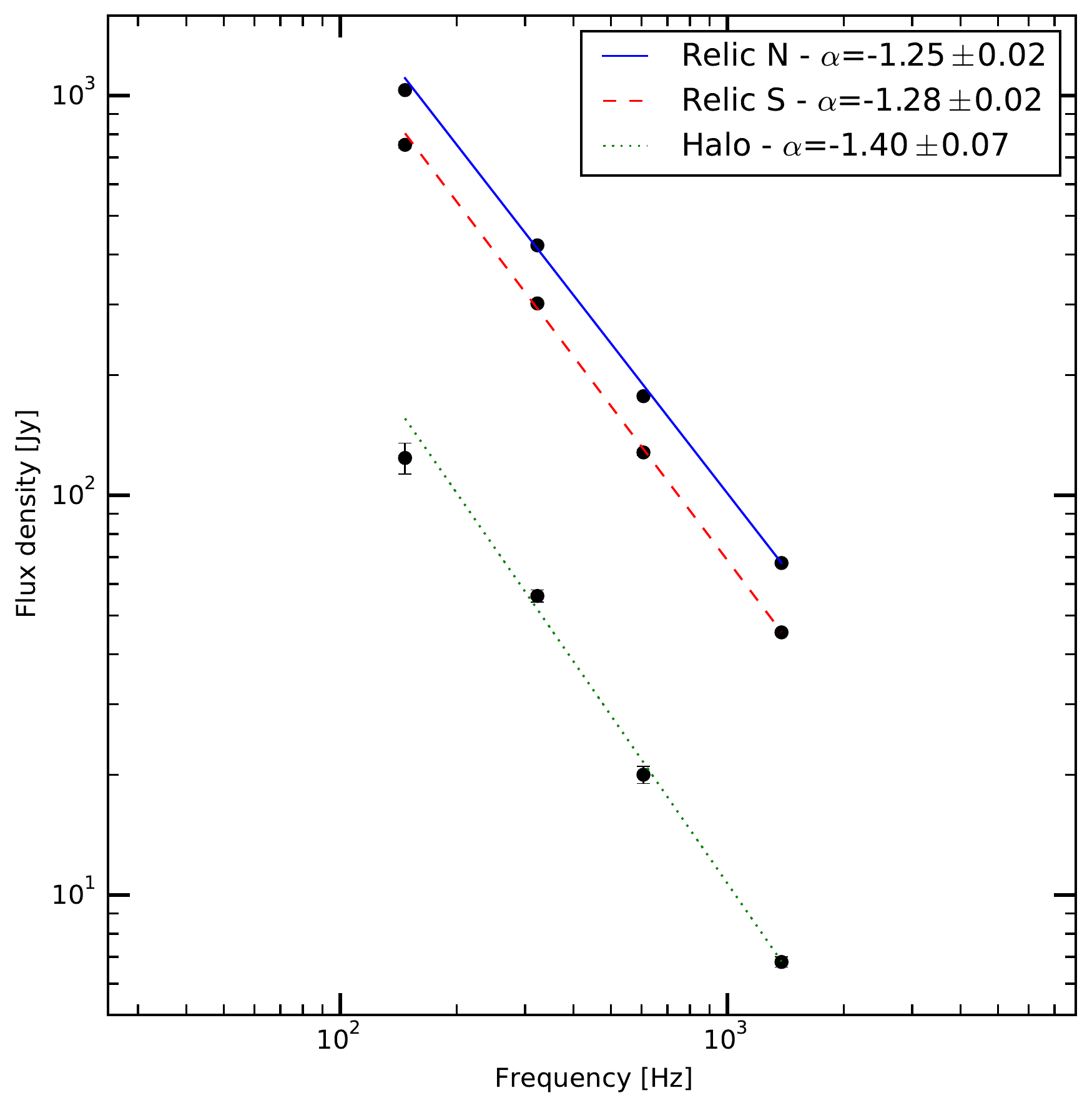}
\caption{Global spectral index analysis for the two radio relic and the radio halo. Solid lines are linear regressions. Error bars are at one $\sigma$ calculated as explained in the text.}\label{fig:globalspidx}
\end{figure}

We extracted global spectral index values by fitting a first order polynomial to the integrated flux densities of the radio halo, the northern relic, and the southern relic (see Fig.~\ref{fig:globalspidx}). The radio halo spectral index value is $\alpha=-1.40\pm0.07$ but data points show a possible curvature of its spectrum as expected by re-acceleration models. The spectral index of the two radio relics is $\alpha=-1.25\pm0.02$ for the northern relic and $\alpha=-1.28\pm0.02$ for the southern relic. Both spectral energy distributions appear flat, as seen for other radio relics as long as the sampling does not span a wide enough range of frequencies \citep[e.g.][]{Stroe2013}. However, as shown by \cite{Trasatti2014}, integrated spectral index values of radio relics may be biased in a number a ways and their interpretation is not straightforward.

\begin{figure}
\centering
\includegraphics[width=\columnwidth]{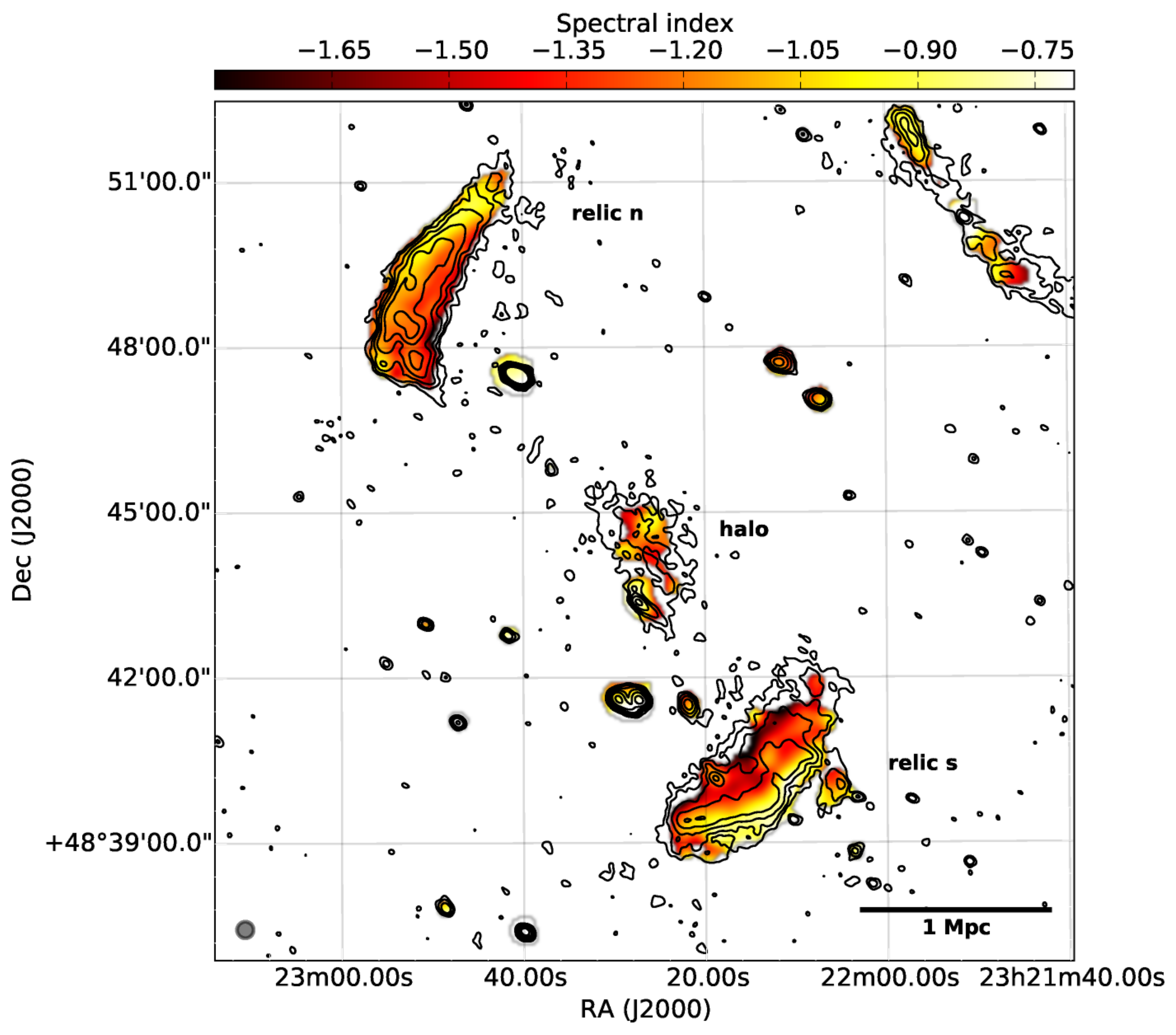}
\caption{Two-frequency spectral index map obtained between the two highest signal to noise ratio maps available: 323 and 1380 MHz. Contour levels as in Fig.~\ref{fig:radio}, radio emission at 323 MHz.}\label{fig:spidx}
\end{figure}

The two-point spectral index map (Fig.~\ref{fig:spidx}) shows a clear gradient (flat $\rightarrow$ steep) going towards the cluster centre along both relic widths. These gradients have been interpreted as a consequence of synchrotron ageing of electrons accelerated by a shock wave on the outer edge of the radio relic \citep[e.g.][]{vanWeeren2010a}, but such interpretation can mask more complex scenarios such as shocks with varying Mach number \citep{Skillman2013}. While for the southern cluster, the steepening happens quite uniformly along the relic extensions, the northern relic's outer edge has a different spectral index between the N and the S side, which may be related to its varying thickness and brightness along its extension. The spectral index of the radio halo is patchy, consistent with previous findings \citep{Orru2007}.

\subsection{Plasma ageing}
\label{sec:ageing}

Given the relatively high brightness of these radio relics we were able to perform a spectral ageing study of the two sources. Compared to radio galaxies, where this type of study has been made several times, only a limited number of radio relics have the required brightness. Moreover, in some cases the complex morphology of the source made the spectral analysis of difficult interpretation \citep[see e.g.][for Abell 2256]{Owen2014}. Only recently extensive spectral age studies have been performed on the ``Sausage'' radio relic \citep{Stroe2013,Stroe2014} and on the ``Toothbrush'' radio relic \citep{VanWeeren2012e}.

The flux calibration errors on all maps were set at 5\%. To ensure recovery of flux on the same spatial scales for the GMRT (147, 323, 607 MHz) and WSRT (1380 MHz) observations we harmonized the $uv$-coverage and convolved the uniform-weighted maps to the same resolution of 22\arcsec. We have used the \texttt{BRATS} package \citep{Harwood2013} to fit the JP model \citep{Jaffe1973} on a pixel-by-pixel basis. The JP model has been shown to be a good assumption for radio relics, or at least not worse than others \citep{Stroe2013}. The radio images of the two relics were loaded separately into \texttt{BRATS}, imposing a $3\sigma$ cut-off on the radio emission based on the off-source rms noise. The northern relic was covered by 196 pixels or an equivalent of almost 20 independent beams. The southern relic emission is fainter and part of the relic cannot match the cut-off criteria in shallower maps. The relic was covered with 115 pixels ($\sim 11$ independent beams).

The JP spectral ageing model has four free parameters: flux normalization, magnetic field $B$, injection index $\alpha_{\rm inj}$, and time since acceleration $t_{\rm age}$. We assumed a uniform magnetic field of $\sim2~\rm\mu G$ \citep[a reasonable assumption][]{Ryu2008a,Bonafede2010,vanWeeren2010a,Iapichino2012} which is able to reproduce the mass-luminosity scaling relation in double radio relics systems \citep{deGasperin2014c}. The shock aligns and likely enhances the magnetic field in the vicinity of the shock front, but the magnetic field condition downstream of the shock front is largely uncertain and the assumption of uniformity is driven by our lack of knowledge. However, \cite{VanWeeren2012e} showed that for the ``Toothbrush'' cluster magnetic field variations do not play a major role in creating spectral trends. Apart from synchrotron radiation, inverse Compton scattering of CMB photons contributes to electrons ageing. \texttt{BRATS} takes the inverse Compton effect into account as an additional magnetic field term $B_{\rm CMB}$ added in quadrature to the original magnetic field $B$. The CMB energy density translated into $B$ field terms is  \citep{Longair2011}:
\begin{equation}
 B_{\rm CMB} = 3.18\, (1+z)^2\ \rm \mu G
\end{equation}
where the redshift is $z=0.335$ and the equivalent magnetic field for the inverse Compton interactions is $B_{\rm CMB} = 5.7\ \rm\mu G$.

The Mach number $\mach$ of a shock is defined as
\begin{equation}
 \mach = v_{\rm shock}/c_s,
\end{equation}
where $c_s$ is the sound speed in the upstream medium. For simple shocks, the $\alpha_{\rm inj}$ values can be used to infer the shock Mach number $\mach$ following \citep{Blandford1987}:
\begin{equation}\label{eq:mach}
 \mach = \sqrt{\frac{2\alpha_{\rm inj}+3}{2\alpha_{\rm inj}-1}}.
\end{equation}

\subsection{Injection index}
\label{sec:injection}

\begin{figure}
\centering
\includegraphics[width=\columnwidth]{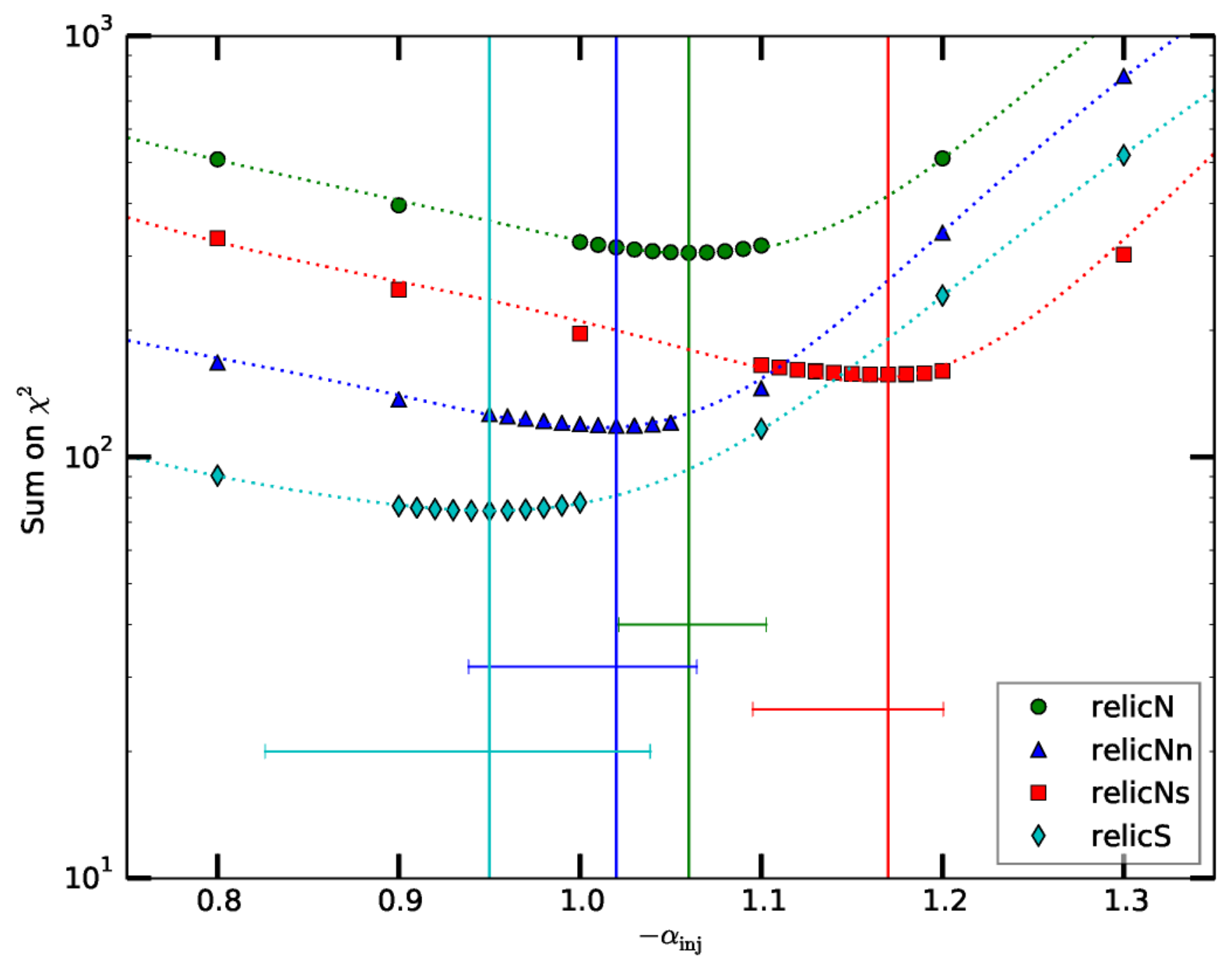}
\caption{Injection index $\alpha_{\rm inj}$ for the entire northern relic (green circles), then N part of the northern relic (blue triangles), the S part of the northern relic (red squares) and the entire southern relic (cyan diamonds). Dotted lines are a 6th order polynomial fit to the data used to find the function minima. Bars in the lower part of the plot show the $1\sigma$ error in the $\alpha_{\rm inj}$ estimation.}\label{fig:injection}
\end{figure}

For each relic, we estimated an injection spectral index $\alpha_{\rm inj}$ by fitting the JP model with a range of $\alpha_{\rm inj}$ and leaving normalization and $t_{\rm age}$ free to vary. We made an initial injection index search between 0.5 and 1.0 in steps of 0.1, followed by a fine search in steps of 0.01. The goodness of fit (sum of $\chi^2$) dependence on the assumed injection index has a single minimum (see Fig.~\ref{fig:injection}), which represents the $\alpha_{\rm inj}$ that best describes the data. We note that it was common practice to estimate the injection index for a relic by looking at the flattest spectral index values close to the shock front or the integrated spectral index value. These approaches are subject to a number of limitations described in \cite{Stroe2014}. Conversely, estimating $\alpha_{\rm inj}$ as described here uses all the available data, minimizing the chances that a noise minimum/maximum in one of the maps can bias our result. As shown in Fig.~\ref{fig:injection} the $\chi^2$ curve is well defined, hence the uncertainty in the injection index can be calculated following \cite{Avni1976} who shows that the $1\sigma$ error is given by a unit change in the $\chi^2$ value from its minimum. As we are considering the source on a pixel-by-pixel basis, the sum of $\chi^2$ is over weighted by a factor of the beam area in pixel $A_{\rm beam}$. Thus, to find the errors, the sum of the $\chi^2$ has to be divided by $A_{\rm beam}$. An important caveat of this approach is the consequence of the limited resolution imposed by the beam in out maps. Mixing the emission from regions with different magnetic fields and/or different spectral ages, moves the spectra closer to power-law shapes, smearing the curvature \citep{VanWeeren2012e}.

\begin{figure*}
\centering
\includegraphics[width=.45\textwidth]{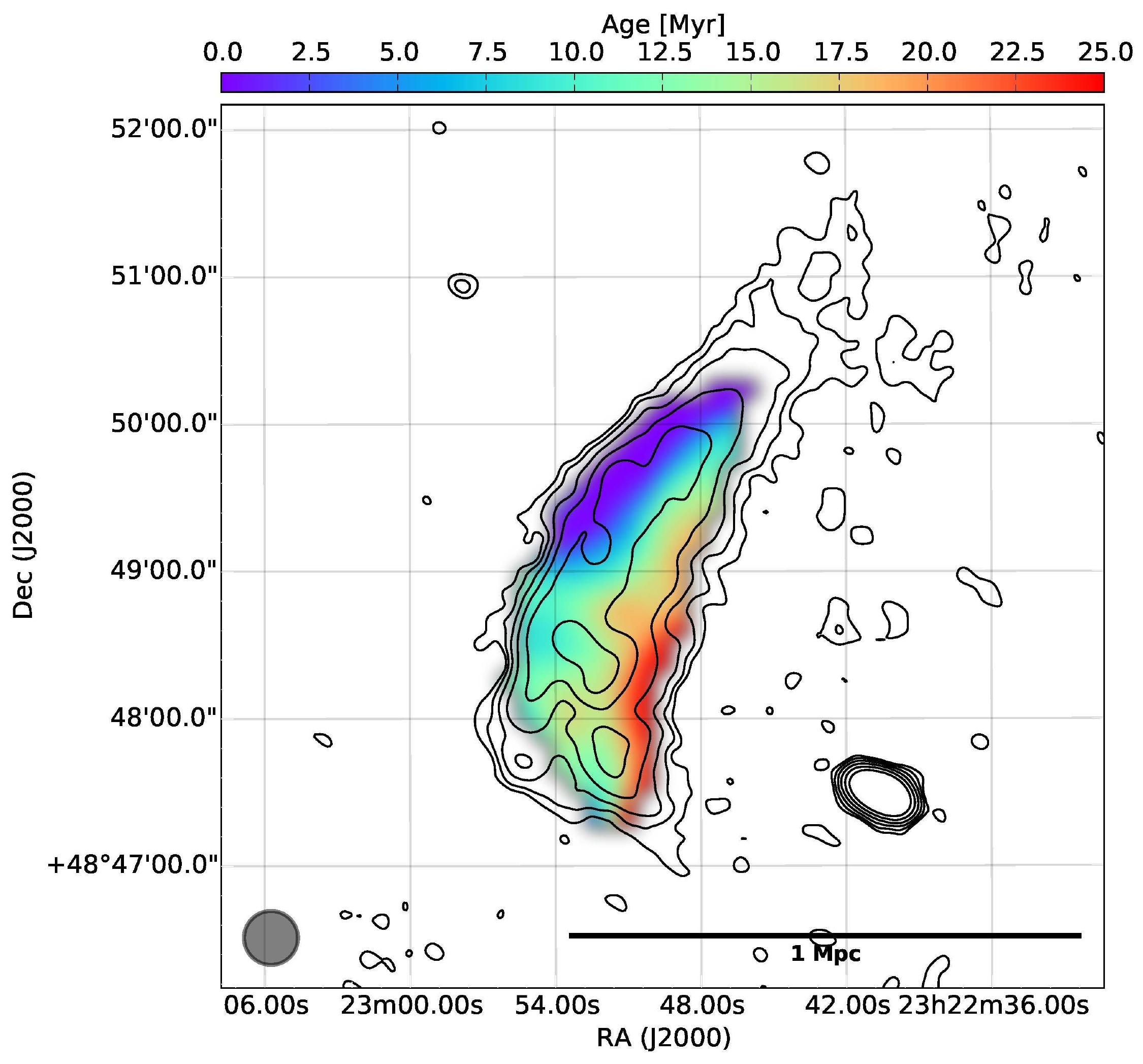}
\includegraphics[width=.45\textwidth]{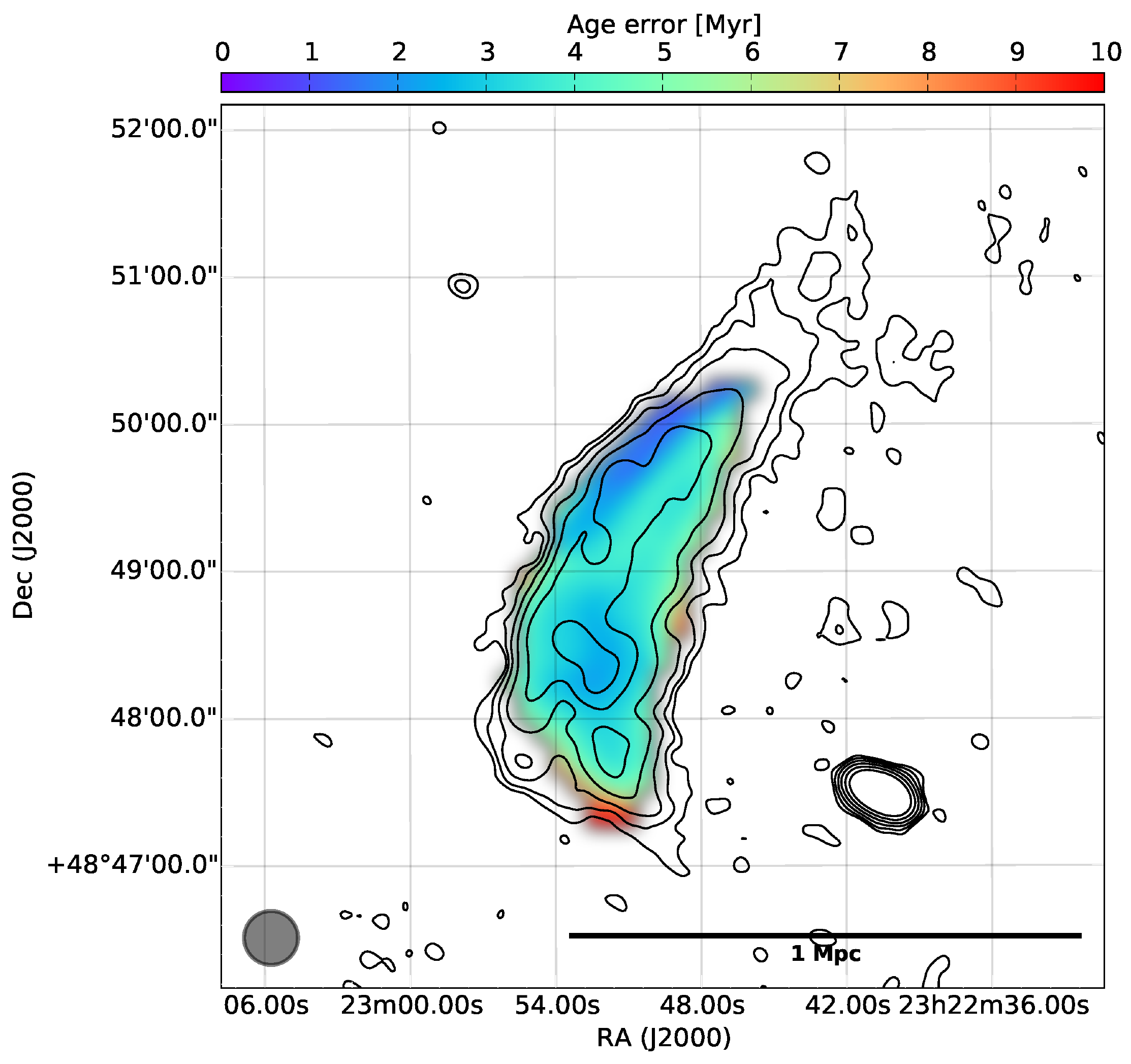}\\
\includegraphics[width=.45\textwidth]{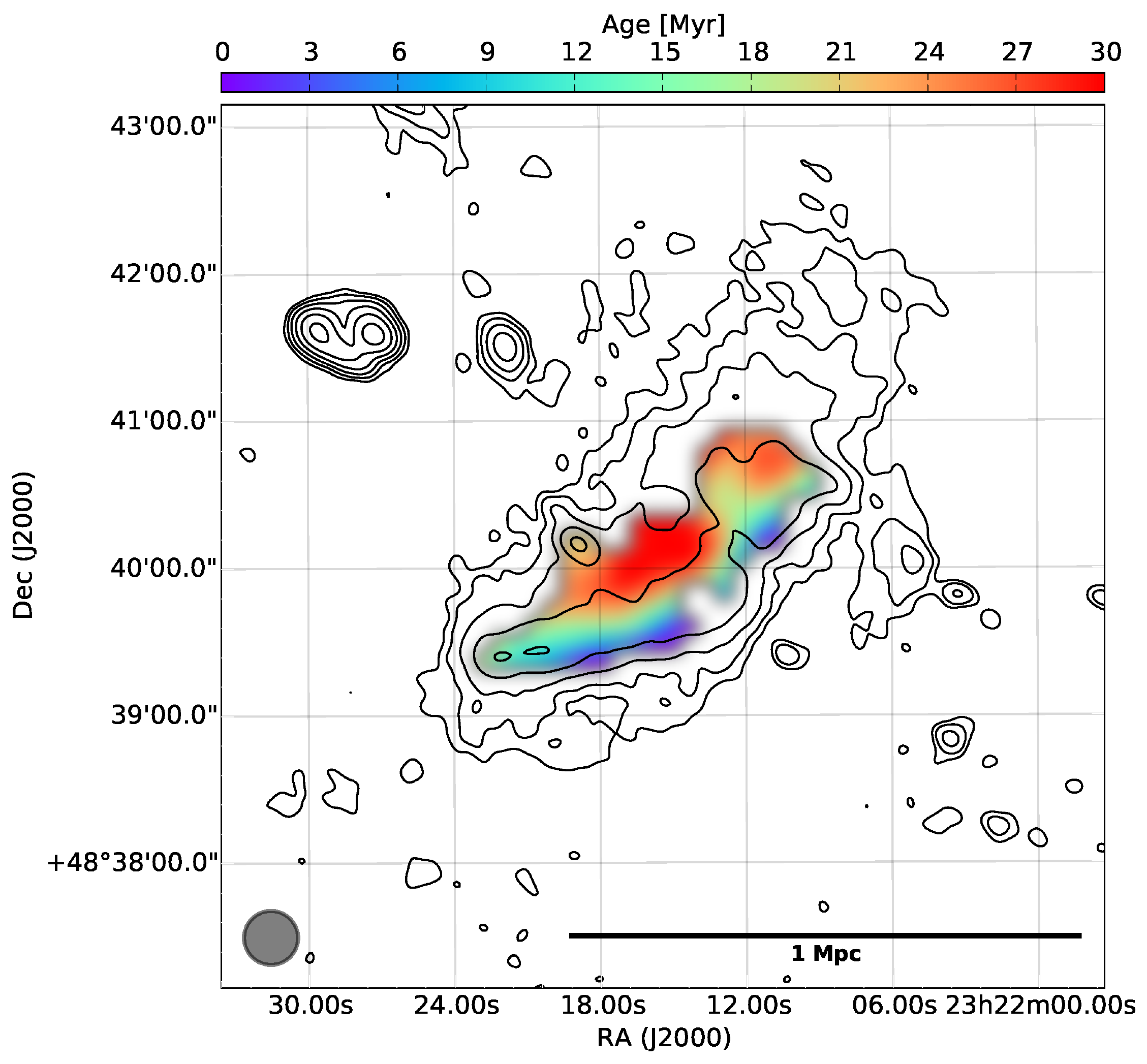}
\includegraphics[width=.45\textwidth]{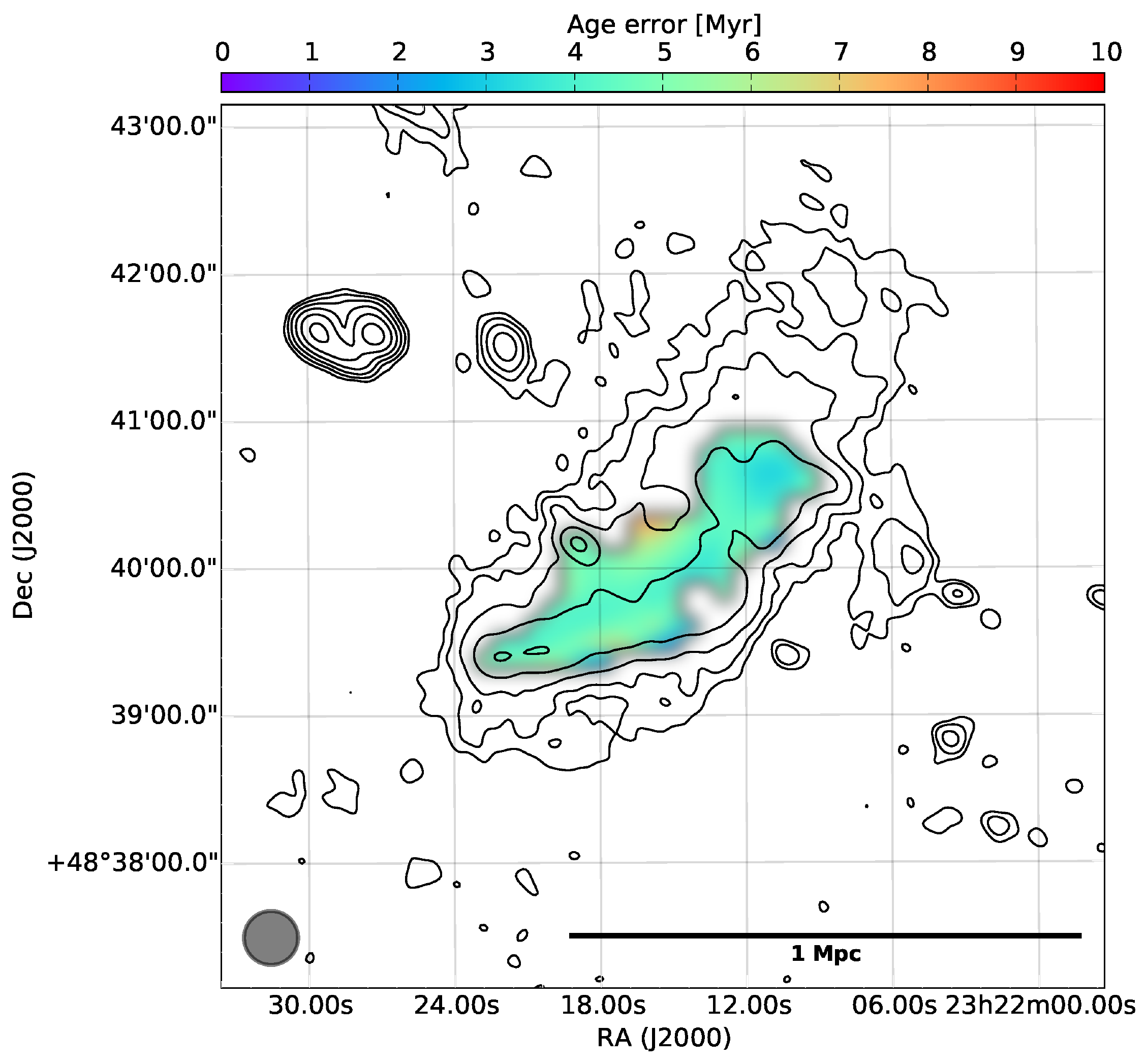}\\
\includegraphics[width=.45\textwidth]{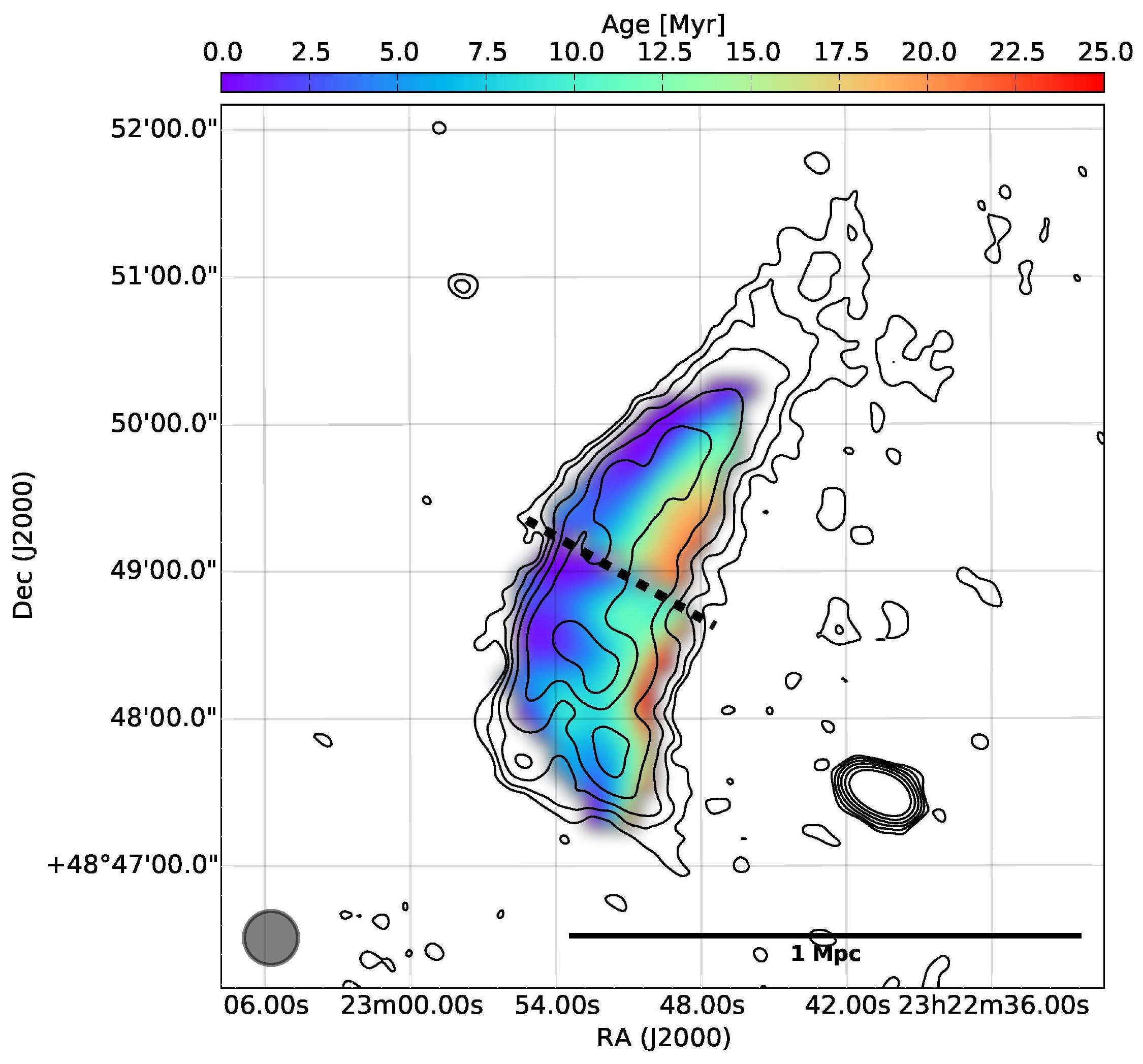}
\includegraphics[width=.45\textwidth]{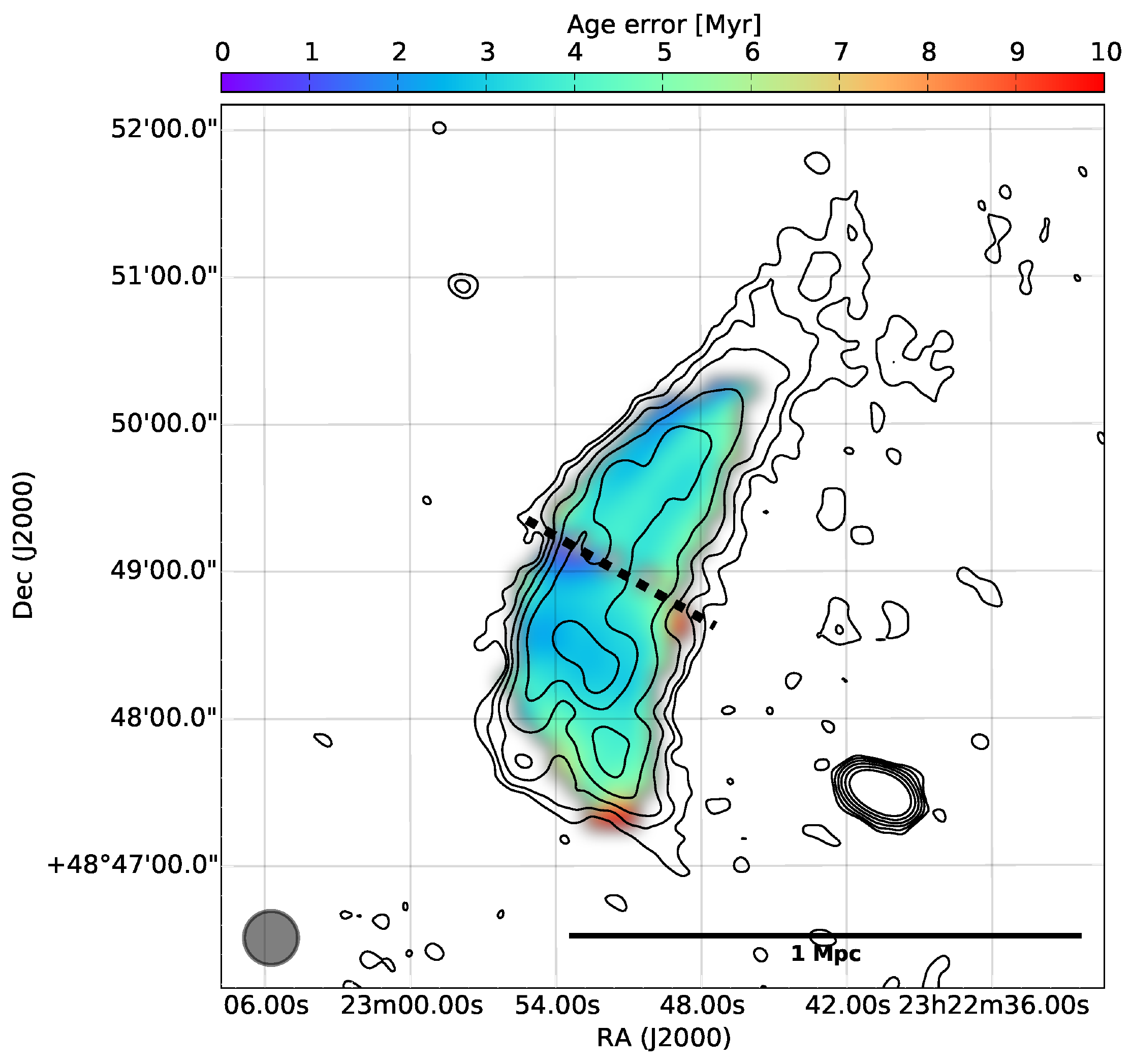}
\caption{Particle age maps based on spectral ageing analysis, with uncertainty maps in the second column. Top and middle rows are for the northern and southern relic respectively; in the bottom row the injection index was fixed at two different values for the region north and south of the dashed lines (see Sec.~\ref{sec:injection}).}\label{fig:age}
\end{figure*}

As a next step, we fixed the injection value to the value we found: $\alpha_{\rm inj,N} = -1.06^{ +0.04 }_{ -0.04 }$ and to $\alpha_{\rm inj,S} = -0.95^{ +0.09 }_{ -0.12 }$ for the northern and southern relic respectively. Finally, we fit a JP model with two degrees of freedom: normalization and $t_{\rm age}$. The resulting maps are in the first two columns of Fig.~\ref{fig:age}. The southern relic displays a quite uniform gradient from 0 to $\sim30$ Myr across the fitted region. For the northern relic the situation is more complex: the region associated with a $t_{\rm age}=0$ (the shock front) appears located only on the northern half of the relic while the aged electrons seem concentrated towards the south. This configuration would be hard to explain with current radio relic theories which point towards a shock present along all the relic extension, although we cannot exclude \textit{ad-hoc} projection effects. However, as noticed in simulations, shocks do not have a uniform Mach number along their extension \citep{Skillman2013}. Therefore, we attempted a separate fit for the $\alpha_{\rm inj}$ in the northern and southern half of the relic. The results are shown in Fig.~\ref{fig:injection}. While the global fit found an $\alpha_{\rm inj,N} =  1.06 ^{ +0.04 }_{ -0.04 }$, separate fits of the two regions returned $\alpha_{\rm inj} =  1.02 ^{ +0.04 }_{ -0.08 }$ and $\alpha_{\rm inj} =  1.17 ^{ +0.03 }_{ -0.07 }$ for the northern and the southern half, respectively. The difference between the two injection indices is statistically significant. Fitting again for $t_{\rm age}$, this time fixing $\alpha_{\rm inj}$ to different values for the two relic regions, produced the map shown in the third column of Fig.~\ref{fig:age}. The particle acceleration ($t_{\rm age}=0$) happens now all along the external relic edge and the ageing happens along the relic shorter axis. It appears that some gradient is still present, but it would likely disappear if a continuum of $\alpha_{\rm inj}$ was used.

With the important assumption that the injection index -- Mach number relation derived from DSA (Eq.~\ref{eq:mach}) is correct, we estimated for the southern relic a Mach number of $\mach=2.33^{+0.19}_{-0.26}$, while the northern relic Mach number goes from $\mach \sim 2.20^{+0.07}_{-0.14}$ of the north part down to $\mach \sim 2.00^{+0.03}_{-0.08}$ of the southern region. this can be the first observational evidence of varying Mach number along a cluster merger shock.

\section{Discussion}
\label{sec:discussion}

\subsection{Injection index estimation}
An alternative approach to estimate the injection index is to use the average spectral index value of the shock region. However, this approach has some limitation as the definition of the region is arbitrary. In our case we used a region which covers the entire external edge of the relics, with a thickness of one beam. The spectral indices obtained in this way are: $\alpha_{\rm inj,N} \sim -1.1\pm0.1$ and $\alpha_{\rm inj,S} \sim -0.9\pm0.1$. Meanwhile, using the integrated spectral index approach we would have underestimated the injection index, obtaining $\alpha_{\rm inj,N} = -0.75\pm0.01$ and $\alpha_{\rm inj,S} = -0.78\pm0.01$. Many works in the literature use one of these two techniques to estimate $\alpha_{\rm inj}$ \citep[see e.g.][for MACSJ1149.5+2223, MACSJ1752.0+4440, Abell 521, Abell 2744, Abell 754, CIZA J2242.8+5301 (``Sausage''), 1RXS J0603.3+4214 (``Toothbrush'')]{Bonafede2012a, Giacintucci2008, Orru2007, Macario2011, Stroe2013, vanWeeren2010a, VanWeeren2012e}. In many cases, the inferred Mach number is higher than the Mach number measured from X-ray observations when they are available. For the Sausage relic, for instance, the temperature jump in X-ray observations indicates a Mach number of $\mach \approx 3.2$ \citep{Akamatsu2013}, lower than the radio Mach number of $\mach \approx 4.6$ derived by \cite{vanWeeren2010a}. On the other hand, for this cluster, by using the same method described in this paper, \cite{Stroe2014} obtained a compatible injection index $\alpha_{\rm inj} \approx 0.77$ corresponding to $\mach \approx 2.9$. For the Toothbrush relic, the temperature and density discontinuities suggest $\mach \approx 2$ \citep{Ogrean2013}, while from radio observation the Mach number is $\mach = 3.3 - 4.6$. Recently \cite{Hong2015} showed using cosmological hydrodynamical simulations that this discrepancy might be due to an observational bias, as radio observations tend to pick up stronger shocks compared to X-ray data. Deep X-ray observation of \target{} may unveil the presence of temperature discontinuity and will allow an independent estimation of the Mach number.

In this paper we showed that with high-quality data we can find different injection indices along a single radio relic. Therefore, assuming the flatter spectral index region of the relic as representative of the true injection index, would bias the estimation towards the region with the strongest Mach number (flatter spectra). Furthermore, a noise minimum/maximum in one of the maps might also bias the estimation towards flatter spectra \citep{Stroe2014}. In an ideal case, the integrated spectra would be the best indicator as it would be marginally influenced by projection effects and uses all available data. However, the simple relation $\alpha_{\rm integrated} = \alpha_{inj} + 0.5$ does not hold in the case of evolving shocks \citep{Kang2015}. Furthermore, integrated spectral energy distributions of radio relics are not power laws when a large frequency range is considered \citep{Trasatti2014, Stroe2014a}. This implies that the simple idea of a superimposition of spectra with an exponential cut-off (the so called continuous injection model) is not valid. Explanations for this include a non-constant magnetic field, a non-uniform Mach number, and/or that the electron acceleration process in low-Mach number shocks is not fully understood.

\begin{figure*}
\centering
\includegraphics[width=.96\columnwidth]{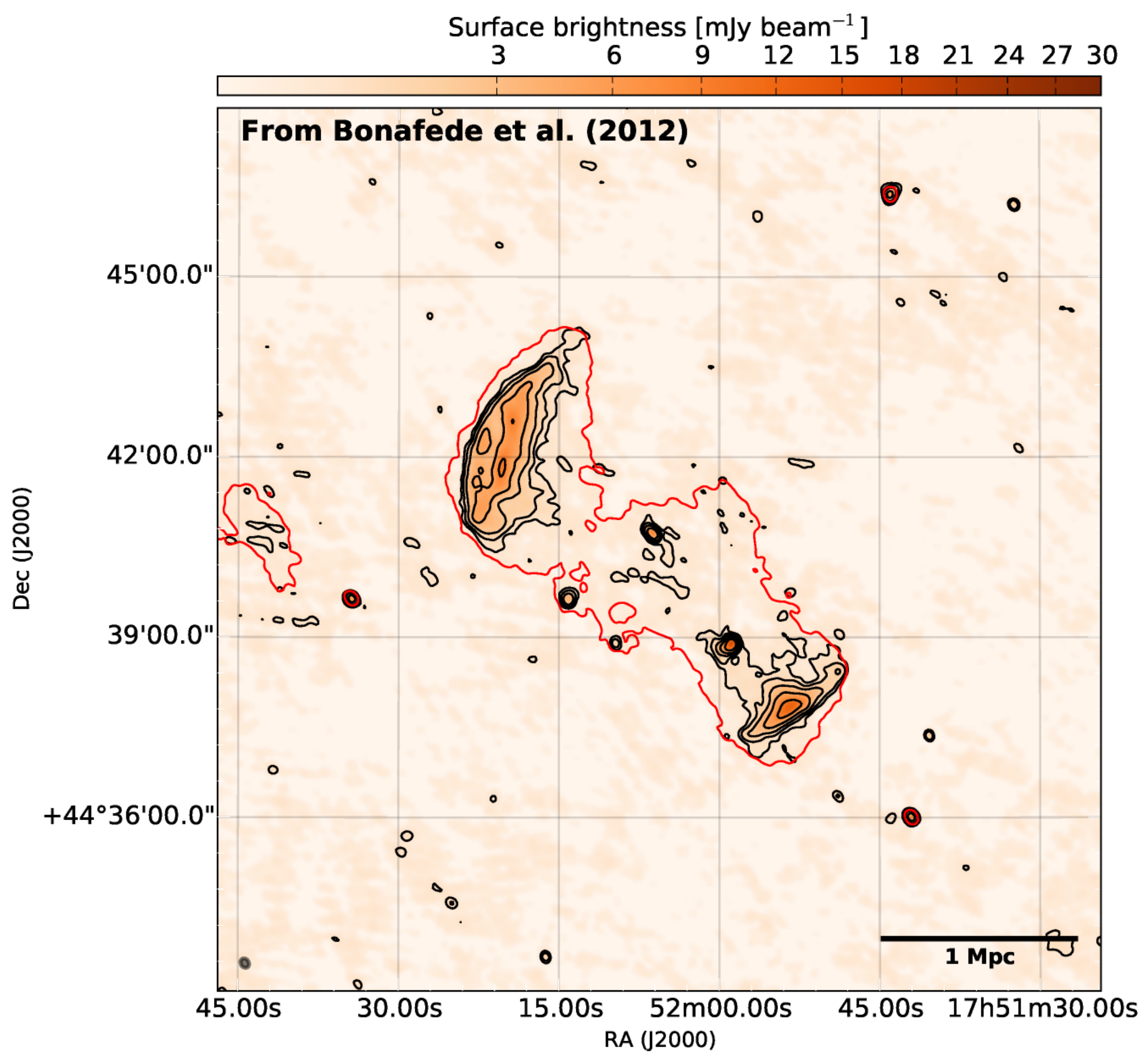}
\includegraphics[width=\columnwidth]{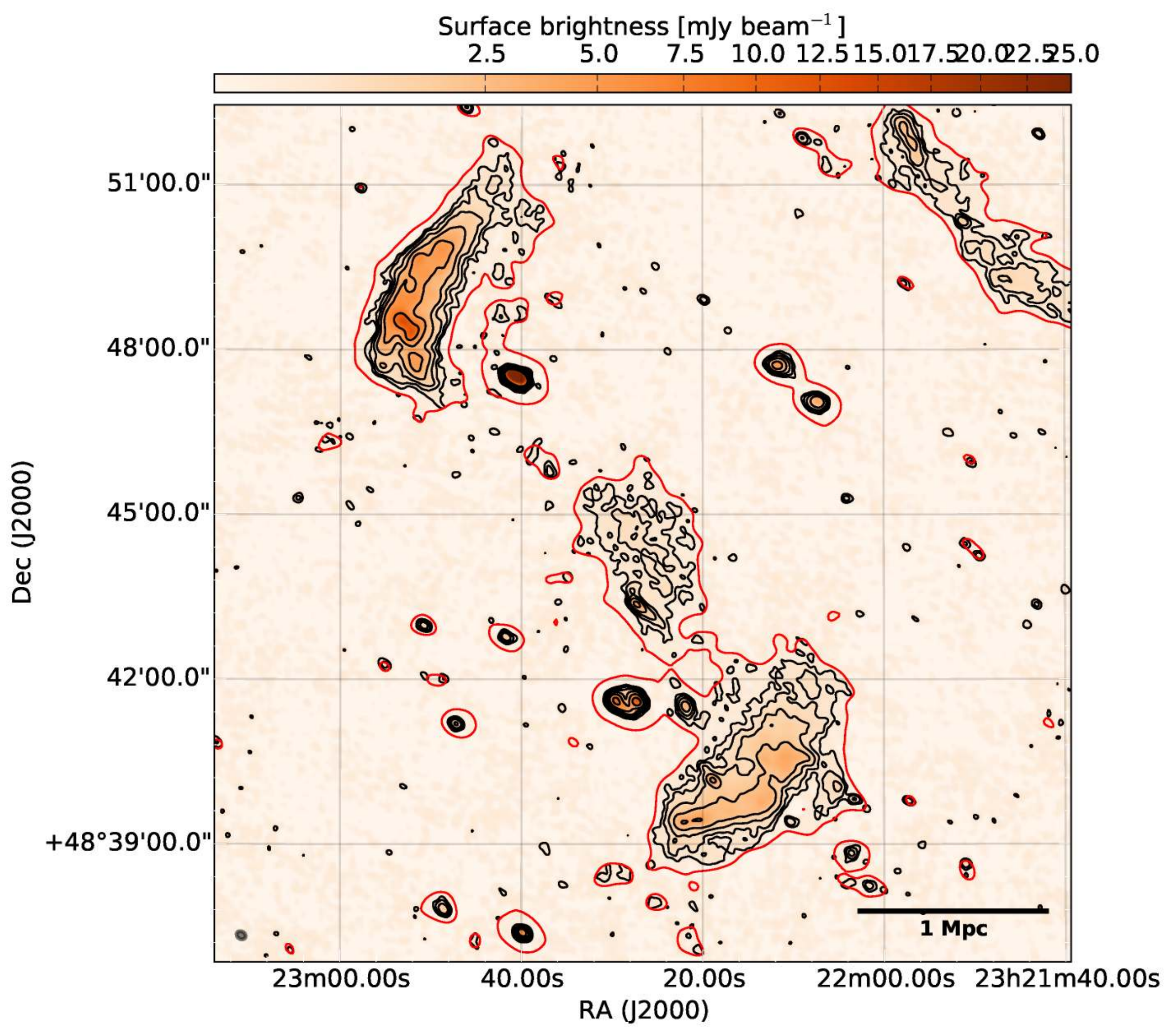}
\caption{Comparison between the image at 323 MHz of MACSJ1752.0+4440 \citep[left;][]{Bonafede2012a} and of \target{} (right). Contours in both cases are at $(1,2,4,8,16,32) \times 3\sigma$ with $\sigma = 180$~\mujybeam and $\sigma = 86$~\mujybeam respectively. Red contour is the $3\sigma$ level for the low-resolution version of the maps.}\label{fig:MACS}
\end{figure*}

\subsection{The sibling: MACS\,J1752.0+4440}
\begin{figure}
\centering
\includegraphics[width=\columnwidth]{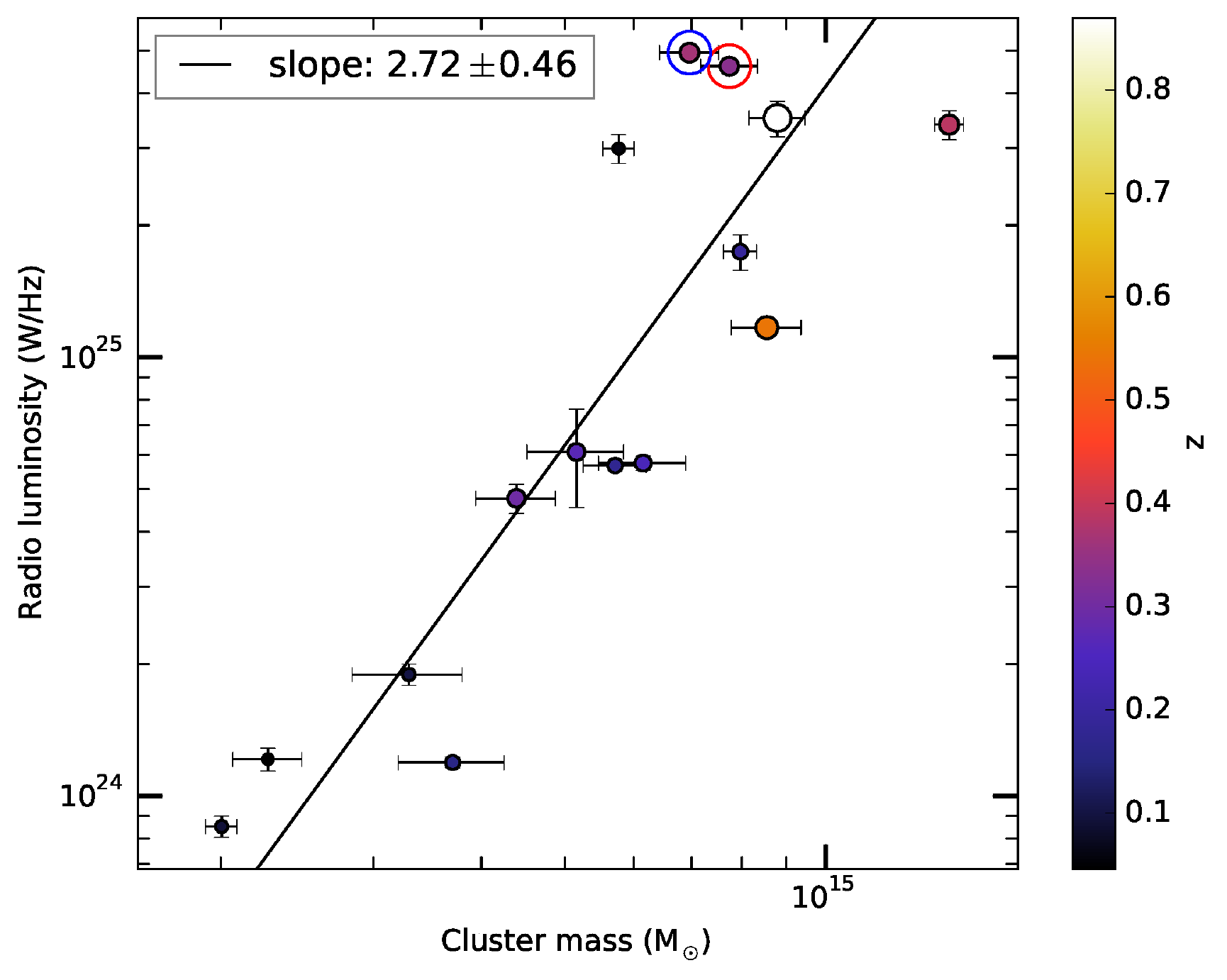}
\caption{Cluster mass ($M_{500}$) versus the hosted double radio relic system luminosity. The correlation in this figure is discussed in \citet{deGasperin2014c}, the red circle is \target{} while the blue circle is MACSJ\,1752.0+4440.}\label{fig:coup_M-Lr-z}
\end{figure}

\target{} appears from many points of view very similar to MACS\,J1752.0+4440 \citep{Bonafede2012a,vanWeeren2012b}. Both clusters are at a similar redshift ($z=0.335$ and $z=0.366$), they are well-defined double radio relic + halo systems, and their radio relics are the two most luminous ever discovered (see Fig.~\ref{fig:coup_M-Lr-z}) with a very similar luminosity. The two systems are also hosted in clusters with comparable SZ-derived masses: $M_{500}=(7.7 \pm 0.6) \times 10^{14} $~\Msun{} and $M_{500}=(6.7\pm0.4) \times 10^{14}$~\Msun, respectively. The two radio relic systems are also impressively similar from the morphological point of view. Both have a relic brighter than the other and they are not perfectly symmetric with respect to the cluster centre. Furthermore, both systems show similar degrees of polarization ($\sim 10\%$) with peaks of 30--40\% on the two northern relics. The orientation of the magnetic field also looks compatible.

A major difference is the relative distance between the relics. In the case of \target{} this is $\sim3100$ kpc, while for MACSJ1752.0+4440 it is around $2/3$ of this, which may imply a more recent merger. Assuming an average shock Mach number of 2 for both clusters as derived from radio observations, we can infer the upstream temperature from the Rankine-Hugoniot relations for a $\Gamma=5/3$ gas to be
\begin{equation}
 \frac{T_{\rm d}}{T_{\rm u}} = \left(\frac{5}{4}\mach^{2}-\frac{1}{4}\right) \cdot \left(\frac{\frac{1}{3}\mach^{2}+1}{\frac{4}{3}\mach^{2}}\right).
\end{equation}
Here $T_{\rm u}$ is the upstream temperature and $T_{\rm d}$ is the downstream temperature which we equate to the mean cluster temperature. The upstream temperature can then be converted into a sound speed $c_s=1480\sqrt{T/(10^8\rm K)}$~\kms, yielding $c_s\sim870$~\kms for \target{} and $c_s\sim880$~\kms for MACS\,J1752.0+4440. Finally, combining this information with the relic distance we can calculate the collision time\footnote{With ``collision time'' we mean the time since the two cluster cores have reached their minimum separation. This time is $\approx 1$~Gyr shorter than the time since the centre of mass of the subcluster crosses the virial radius of the main cluster \citep{Tormen2004}.} to be $\sim870$ Myr ago for \target{} and $\sim530$ Myr for MACS\,J1752.0+4440. Likely the mean cluster temperature is lower than the post-shock temperature \citep[e.g.][]{Markevitch2006}, so these ages should be considered upper limits.

The two relics in \target{} are similar in dimensions, while the relics in MACS\,J1752.0+4440 are asymmetric in power with the flux ratio between the two relics being larger in MACS\,J1752.0+4440. The most striking difference between the two system is the radio halo which is centrally located and disconnected from the relics in \target{}, while in MACS\,J1752.0+4440 it extends touching both radio relics \citep{vanWeeren2012b}. Possible explanations include:

\begin{enumerate}
 \item The radio halo in MACS\,J1752.0+4440 is at its maximum extension while the halo in \target{} is already in its fading phase, in line with its relatively steep spectral index \citep{Donnert2013}.
 \item A different merger dynamics caused different amounts of turbulent energy to be injected into the ICM.
 \item The radio halo follows the magnetic field which traces the X-ray distribution. The X-ray brightness distribution appears more compact in \target{} while it is elongated in MACS\,J1752.0+4440 \citep{vanWeeren2012b}, however this will need to be confirmed by Chandra observations. 
 \item The spatial distribution of seed electrons could differ in the two clusters. We do not know yet how CRs accumulate in clusters over time, but they live for $\sim 1$~Gyr and at very low levels (0.1\%) are expected to be injected from supernova/AGN feedback \citep{Brunetti2014}. Hence, this explanation only shift the problem towards the seed particles.
\end{enumerate}

If indeed radio relics behave self-similarly, i.e. similar merger configurations produce similar radio relics, the origin of radio relics should be generic. If radio relics require shock re-acceleration \citep[e.g.][]{Bonafede2014a}, the pre-existing radio dark CR populations also needs to be similar across clusters. This would make it unlikely that the fortuitous co-location with a source of seed CR --as an aged radio galaxy lobe-- is a requirement for the presence of a radio relic.

\section{Conclusions}
\label{sec:conclusions}

We reported the discovery of a new double radio relic + radio halo system in \target{}. The system is one of the (now) 16 systems with two radio relics. Among these systems only six others also show a clear radio halo. Our main results are summarized here:
\begin{itemize}
 \item \target{} hosts the second most powerful radio relic system known, with a total power of $L_{\rm 1.4 GHz} = (44.6 \pm 0.3) \times 10^{24}$ W/Hz. The cluster also hosts a radio halo of power $L_{\rm 1.4 GHz} = (2.8 \pm 0.1) \times 10^{24}$ W/Hz. Both, radio relic and halo emission agree with luminosity-mass scaling relations found by \citep{deGasperin2014c} and \citep{Cassano2013}, respectively.
 \item We measured up to $\sim10-30\%$ polarized emission from both the northern and southern relic. The magnetic field in the northern relic appears well aligned with the relic extension.
 \item We measured the injection index of the radio spectrum for both relics. We found that the northern relic local spectra are better described by two different injection indices, $\alpha_{\rm inj} =  1.02 ^{ +0.04 }_{ -0.08 }$ and $\alpha_{\rm inj} =  1.17 ^{ +0.03 }_{ -0.07 }$ for the northern and the southern half, respectively.
 \item The Mach number derived from the injection index according to simple DSA theory (see Eq.~\ref{eq:mach}) has been measured to be $\mach = 2.33^{+0.19}_{-0.26}$ for the southern relic, while the northern relic Mach number goes from $\mach = 2.20^{+0.07}_{-0.14}$ in the north down to $\mach = 2.00^{+0.03}_{-0.08}$ in the southern region. This indicates a gradient in the Mach number along the shock which produced the northern relic, as predicted by simulations \cite[see e.g.][]{Skillman2013}. The Mach number across the relic varies by $\sim 10\%$.
 \item We note the interesting similarity between MACS\,J1752.0+4440 and \target{}, the first and second most luminous radio relic systems known. They are at similar redshift, have similar host masses, and have similar radio luminosity and relic relative position and morphology. An important difference between the two systems is the relic distance, MACS\,J1752.0+4440 being $2/3$ of \target{}, and the radio halo which connects the two relics in MACS\,J1752.0+4440 \citep{Bonafede2012a,vanWeeren2012b}, while in \target{} is confined in the central part of the cluster.
\end{itemize}

\appendix
\section{GMRT imaging using MSMF algorithm}
\label{sec:msmf}

The novel multi-scale multi-frequency deconvolution algorithm included in latest CASA releases, models the wide-band sky brightness distribution as a combination of both spatial and spectral functions \citep{Rau2011}. The algorithms is therefore capable to perform radio image reconstruction on wide bandwidth observations or on observations carried out at multiple frequencies. The output of the process is a monochromatic brightness distribution map at a chosen frequency and a map of spectral indices extracted from the Taylor series used to model the spectra of the flux components. We jointly deconvolved three GMRT datasets at 147, 323 and 607 MHz, modelling the source spectra with a fist order polynomial. To our knowledge this is the first time this approach is used on GMRT data.

\begin{figure}
\centering
\includegraphics[width=\columnwidth]{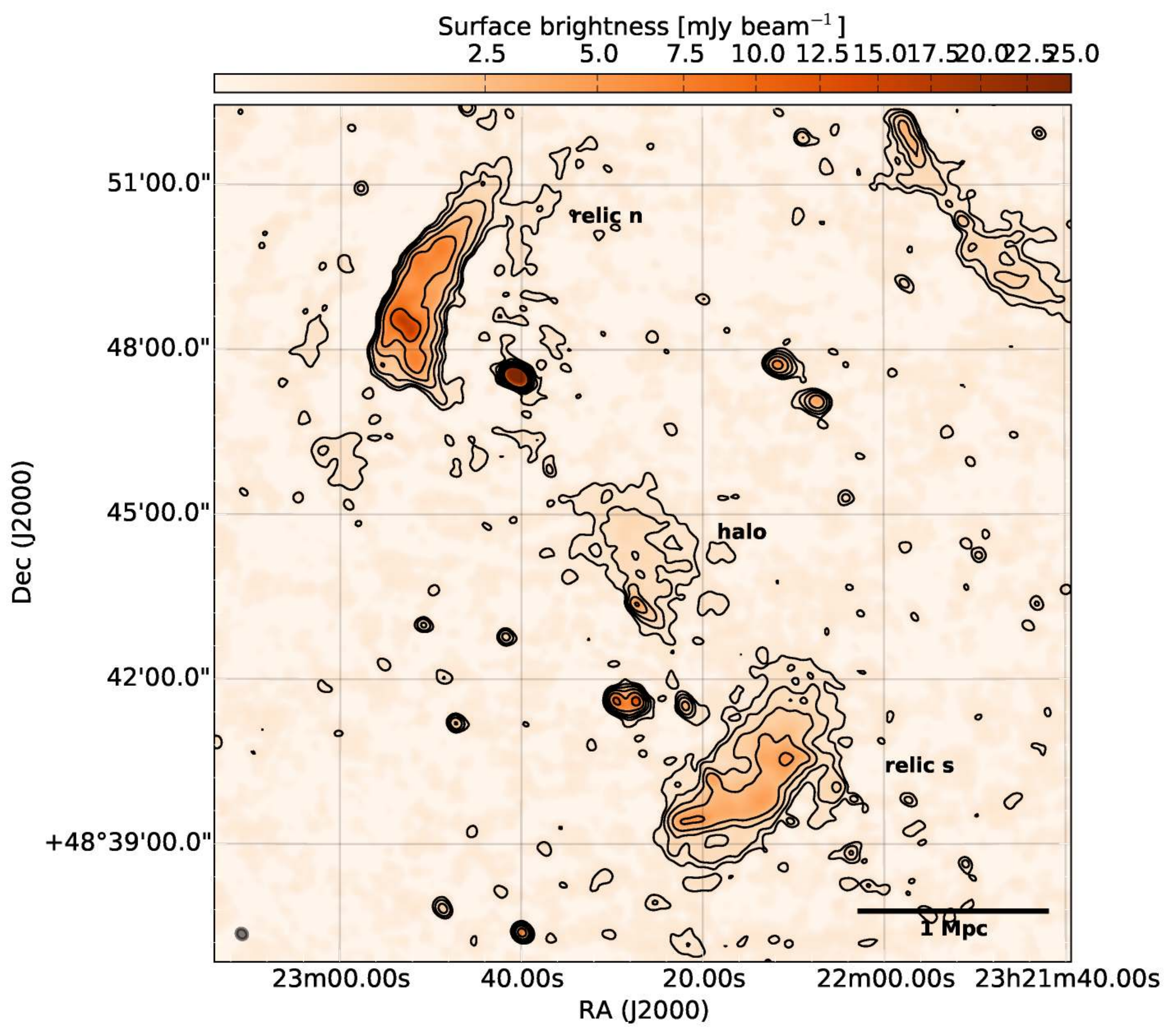}
\caption{Total intensity map obtained using the MSMF algorithm as described in Sec.~\ref{sec:msmf}. The reference frequency is 380 MHz and beam size is \beam{12}{11}. Contours are at $\left(1,2,4,8,16,32\right)\times3 \sigma$ with $\sigma = 100$~\mujybeam.}\label{fig:GMRT-MSMF}
\end{figure}

\begin{figure}
\centering
\includegraphics[width=\columnwidth]{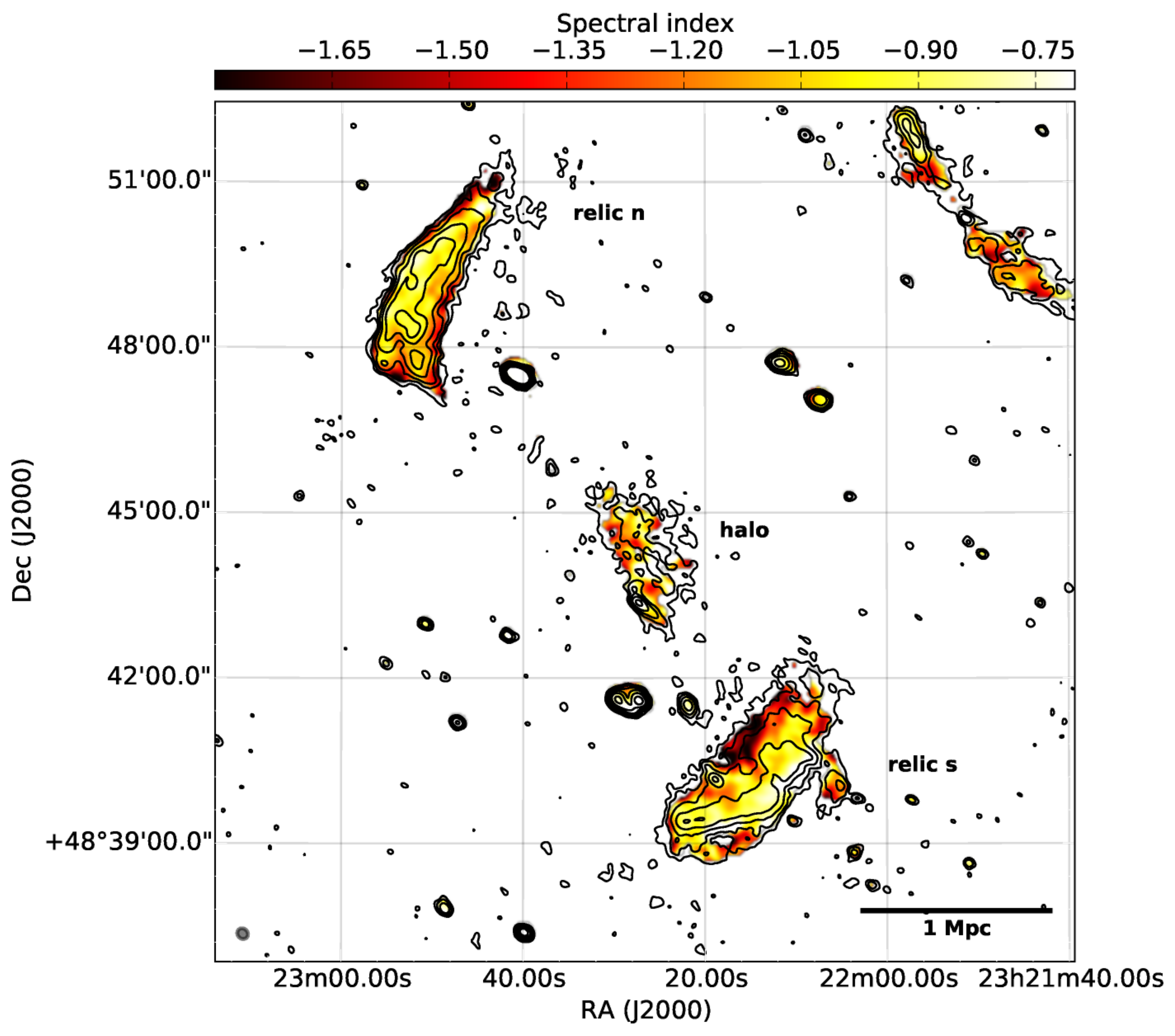}
\caption{Spectral index map obtained using the MSMF algorithm as described in Sec.~\ref{sec:msmf}. Blanked pixels had an error higher than 0.5. Contour levels as in Fig.~\ref{fig:radio}, radio emission at 323 MHz.}\label{fig:GMRT-MSMF-spidx}
\end{figure}

We show the outcome in Fig.~\ref{fig:GMRT-MSMF}, where we produced an image at the intermediate frequency of 380~MHz. The image has a comparable resolution with the 323~MHz image (Fig.~\ref{fig:radio}). Despite the slightly higher noise, the extended emission appears better reconstructed due to a better coverage of the $uv$-plane. We also obtained the spectral index map (Fig.~\ref{fig:GMRT-MSMF-spidx}) which appears qualitatively similar to the 2-point spectral index map shown in Fig.~\ref{fig:spidx}. The 2-point spectral index map was computed using the two highest signal-to-noise ratio maps (323 and 1380 MHz). It is not surprising that sampling higher frequencies than in the MSMF case, the spectral index values are on average $\sim 0.1$ steeper for the relic and $\sim 0.2$ for the halo which has a more curved spectrum.

\section*{Acknowledgements}
We thank Jeremy Harwood for the support in the use of the \texttt{BRATS} package and Julius Donnert for the useful discussions.

AB and MB acknowledge support by the research group FOR 1254 funded by the Deutsche Forschungsgemeinschaft.

R.J.W. is supported by NASA through the Einstein Postdoctoral
grant number PF2-130104 awarded by the Chandra X-ray Center, which is
operated by the Smithsonian Astrophysical Observatory for NASA under
contract NAS8-03060. 

Part of this work performed under the auspices of the U.S. DOE by LLNL under Contract DE-AC52-07NA27344 (LLNL-JRNL-671211).

We would like to thank the staff of the GMRT that made these observations possible. GMRT is run by the National Centre for Radio Astrophysics of the Tata Institute of Fundamental Research.

The Westerbork Synthesis Radio Telescope is operated by the ASTRON (Netherlands Institute for Radio Astronomy) with support from the Netherlands Foundation for Scientific Research (NWO).

Some of the data presented herein were obtained at the W.M. Keck Observatory, which is operated as a scientific partnership among the California Institute of Technology, the University of California and the National Aeronautics and Space Administration.

Funding for the DEEP2/DEIMOS pipelines has been provided by NSF grant AST-0071048. The DEIMOS spectrograph was funded by grants from CARA (Keck Observatory) and UCO/Lick Observatory, a NSF Facilities and Infrastructure grant (ARI92-14621), the Center for Particle Astrophysics, and by gifts from Sun Microsystems and the Quantum Corporation.

This research has made use of SAOImage DS9, developed by Smithsonian Astrophysical Observatory.



This research has made use of the NASA/IPAC Extragalactic Database (NED) which is operated by the Jet Propulsion Laboratory, California Institute of Technology, under contract with the National Aeronautics and Space Administration.

This research has made use of NASA's Astrophysics Data System.

\bibliographystyle{mn2e}
\bibliography{PSZ1G108.18-11.53}
\bsp

\label{lastpage}

\end{document}